\documentclass[preprint,3p,times,11pt]{elsarticle}

\usepackage{lineno,hyperref}
\usepackage[section]{placeins}
\usepackage{graphicx}
\usepackage{mwe}
\usepackage{subfig}

\usepackage{amssymb}
\usepackage{amsmath}
\usepackage{bbm}
\usepackage{natbib}
\usepackage[svgnames]{xcolor}

\modulolinenumbers[1]

\journal{Petroleum}

\begin{document}

\begin{frontmatter}

\title{Machine learning for recovery factor estimation of an oil reservoir: a tool for de-risking at a hydrocarbon asset evaluation}

\author[skoltech]{Ivan Makhotin}
\ead{Ivan.Makhotin@skoltech.ru}
\author[skoltech]{Denis Orlov}
\ead{D.Orlov@skoltech.ru}
\author[skoltech]{Dmitry Koroteev}
\ead{D.Koroteev@skoltech.ru}
\author[skoltech]{Evgeny Burnaev}
\ead{E.Burnaev@skoltech.ru}
\author[zn]{Aram Karapetyan}
\ead{AKarapetyan@nestro.ru}
\author[zn]{Dmitry Antonenko}
\ead{DAntonenko@nestro.ru}

\address[skoltech]{Skolkovo Institute of Science and Technology, Moscow, Russia}
\address[zn]{JSC Zarubezhneft, Moscow, Russia}

\begin{abstract}
Well known oil recovery factor estimation techniques such as analogy, volumetric calculations, material balance, decline curve analysis, hydrodynamic simulations have certain limitations. Those techniques are time-consuming, require specific data and expert knowledge. Besides, though uncertainty estimation is highly desirable for this problem, the methods above do not include this by default. In this work, we present a data-driven technique for oil recovery factor (limited to water flooding) estimation using reservoir parameters and representative statistics. We apply advanced machine learning methods to historical worldwide oilfields datasets (more than 2000 oil reservoirs). The data-driven model might be used as a general tool for rapid and completely objective estimation of the oil recovery factor. In addition, it includes the ability to work with partial input data and to estimate the prediction interval of the oil recovery factor. We perform the evaluation in terms of accuracy and prediction intervals coverage for several tree-based machine learning techniques in application to the following two cases: (1) using parameters only related to geometry, geology, transport, storage and fluid properties, (2) using an extended set of parameters including development and production data. For both cases model proved itself to be robust and reliable. We conclude that the proposed data-driven approach overcomes several limitations of the traditional methods and is suitable for rapid, reliable and objective estimation of oil recovery factor for hydrocarbon reservoir.
\end{abstract}


\end{frontmatter}
\newpage
\section{Introduction}

When bidding for a license area for hydrocarbon exploration, operating companies need to evaluate an expected margin as accurate as possible. A significant portion of overall investment into an oilfield is spending to get as much a-priori information about a reservoir as possible. Estimation of expected oil recovery is essential for the asset evaluation and further field development planning. Oil recovery factor is critically affected by characteristics of the reservoir (geological structure, internal architecture, properties of reservoir rock and fluids) and the specifics of the oilfield development scheme \cite{rui2017quantitative}. There are several methods to estimate oil recovery factor with data collected from seismic surveying or acquisition of a previous surveying data, well logs, petrophysical studies and collection of production profiles. Nowadays, most of the sedimentary basins that contain oil have already been explored. While newly discovered tend to be small. The ability to choose the most cost-effective one among many possible options with varying completeness of data becomes more relevant. That is why it is essential to estimate recoverable reserves of discoveries rapidly and with predictive uncertainty. There are several methods for oil recovery estimation at early stages of the oilfield  exploration (sometimes referred to as greenfield). These methods could be applied at stages when there are no sufficient amount of production data and detailed hydrodynamic model. Volumetric and analogy are the most famous ones \citep{demirmen2007reserves}. The analogy method requires representative oilfields database and highly depends on reservoir characteristics similarity measure. The main idea of the volumetric method is to estimate original oil in place with geological model that geometrically describes the volume of hydrocarbons in the reservoir. Along with this, oil recovery factor evaluation performing by estimating primary and secondary recovery. The primary recovery factor is often estimated mainly from predominant drive mechanism identification. The secondary recovery factor is estimated as the product of displacement efficiency and sweep efficiency. These terms are influenced by fluid flow properties, reservoir heterogeneity that may be measured with petrophysical studies and well logs. Both methods require a specific set of data, relatively time-consuming and do not provide predictive uncertainty by default. There are cases when it might be necessary to assess mature oilfield (sometimes referred to as brownfield). High amount of production data or relatively detailed hydrodynamic model allows obtaining an accurate and reliable reserves estimation using decline curve analysis, material balance or numerical hydrodynamic simulations \citep{demirmen2007reserves}. Decline curve analysis refers to reserves estimation based on production measurements such as oil rate and oil-cut. Material balance and numerical hydrodynamic simulations are good in terms of capturing the major physical mechanisms of hydrocarbon filtration through a reservoir rock. However, these methods are relatively time-consuming, require significant efforts and detailed reservoir description to build an accurate model and even greater efforts to conduct uncertainty quantification.

Nowadays, different machine learning techniques are increasingly being applied in the oil and gas industry \cite{li2020applications}. The data-driven approach allows retrieving non-trivial dependencies and building powerful predictive models from historical data. Several studies are demonstrating empirical relationships between available parameters at exploration phase and oil recovery factor. \citet{guthrie1955use} obtained linear dependency of recovery factor for water drive reservoirs on its properties. \citet{arps1967statistical} obtained non-linear relationships for water drive and solution gas drive reservoirs using the same data. Recently, there were several attempts to apply machine learning to build recovery factor estimation model.
\citet{sharma2010classification} used Tertiary Oil Recovery Information System (TORIS) as oil reservoirs training set and the Gas Information System (GASIS) as gas reservoirs training set to fit multivariate linear regression. The authors demonstrated high accuracy of the linear model. However, less than $2\%$ of TORIS oil reservoirs were used for training and testing. \citet{mahmoud2019estimation} showed the successful application of the artificial neural networks (ANNs) using descriptions of 130 water drive sandstone reservoirs. \citet{han2018hybrid} demonstrated the application of a model based on support vector machine in combination with the particle swarm optimization (PSO-SVM) technique for oil recovery factor prediction using description of 34 low-permeable reservoirs. \citet{aliyuda2019machine} demonstrated successful application of Support Vector Machine using 93 reservoirs descriptions from the Norwegian Continental Shelf as a dataset.

The objective of this study is to construct and evaluate a general, rapid and robust data-driven model (surrogate model, \cite{GTApprox2016}) for oil recovery factor limited to water flooding) estimation with predictive uncertainty. In \nameref{sec:datasets} section we describe data we use to build training set. Relatively rich available sources of data characterizing more than 2000 oil reservoirs from all over the world allow constructing representative training sample to attain strong generalization ability. \nameref{sec:background} section briefly introduces notions and algorithms from machine learning theory we use. We consider the application of machine learning regression algorithms based on trees ensembles, which are well suited to problem specificity. We also use clustering and dimensionality reduction methods for analysis. \nameref{sec:methodology} section describes evaluation methods and details of both model for pre-production phases and model for post-production phases. Results of numerical experiments are presented in \nameref{sec:results}. In the \nameref{sec:discussion} we interpret our results and compare it with other studies. In the \nameref{sec:conclusion} we briefly describe the main idea of the paper, most important findings, highlight overall significance of the study and state future directions.

\section{Datasets}
\label{sec:datasets}

We consider two datasets. Both contain multi-feature oil reservoirs descriptions. The description includes time-independent characteristics and parameters related to reservoir geometry, geology and petrophysical studies. Datasets also contain a set of parameters measured at some moment during the production phases. Both datasets include expected ultimate recovery (EUR), which were estimated according to the development plans and measurements assuming primary and secondary recovery. Overall, databases provide information about 2500 oil reservoirs all over the world.

\subsection{TORIS dataset}
\begin{figure}[!h]
\centering
    \includegraphics[width=.9\textwidth]{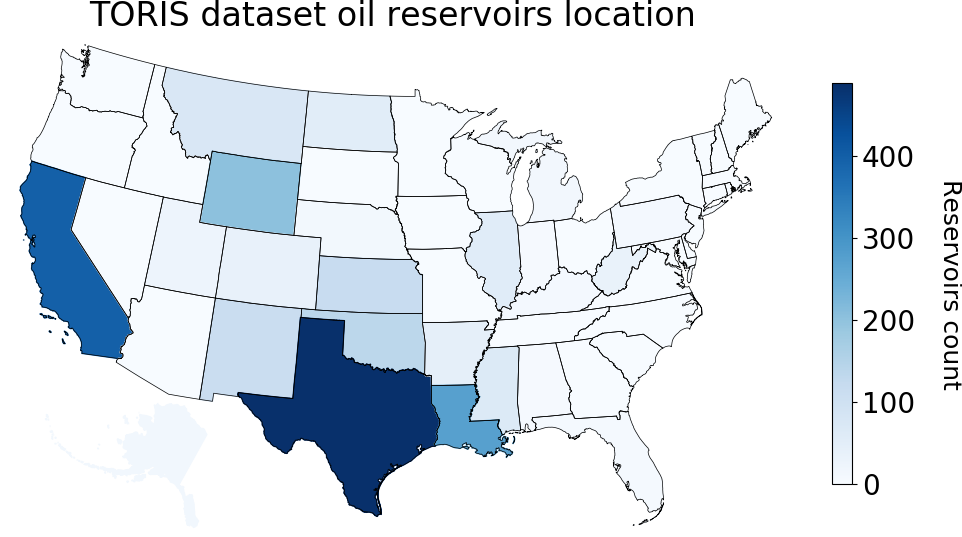}
    \caption{Oil reservoirs location from TORIS database}
    \label{fig:TORIS_map}
\end{figure}

Tertiary Oil Recovery Information System (TORIS) is a data repository, which was originally developed by the National Petroleum Council (NPC) for its 1984 assessment of the USA enhanced oil recovery potential \citep{toris, sharma2010classification}. Dataset contains description of 1381 oil reservoirs from the USA. Number of parameters is 56. Among them 12 categorical and 44 numerical. Data contains incomplete records, 22\% values are missing. Geographical layout is shown in Figure \ref{fig:TORIS_map}.

\FloatBarrier

Only 831 of 1381 reservoirs contain expected ultimate oil recovery factor and can be used for training and evaluation. We group all parameters in the following way:

\begin{itemize}
\item Geometry | Field Acres, Proven Acres, Net Pay, Gross Pay, True Vertical Depth, Reservoir Acres, Reservoir Dip
\item Geology | Lithology, Geologic Age, Fractured-Fault, Shale break of laminations, Major Gas Cap, Deposition System, Diagenetic Overprint, Structural Compartmentalization, Predominant Element of Reservoir Heterogeneity, Trap Type
\item Transport, Storage and Fluid properties | Porosity, Permeability, Oil viscosity, Formation salinity, Clay content, Formation temperature, API Gravity
\item Saturations, Ratios and Pressures | Initial \& Current oil saturations, Initial \& Current water saturations, Initial \& Current gas saturations, Initial \& Current oil formation volume factor, Initial \& Current formation pressure, Initial \& Current producing GOR
\item Development and Production | Well Spacing, Production/Injection wells count, Swept Zone oil saturation (Residual to water), Injection water salinity, Dykstra-Parsons Coefficient, Current injection rate, Original oil in place, Production rate, Cumulative oil production, First stage oil recovery factor, Second stage oil recovery factor
\item Location | State, Formation Name
\end{itemize}

\subsection{Proprietary dataset}

\begin{figure}[!h]
  \centering
    \includegraphics[width=1.\textwidth]{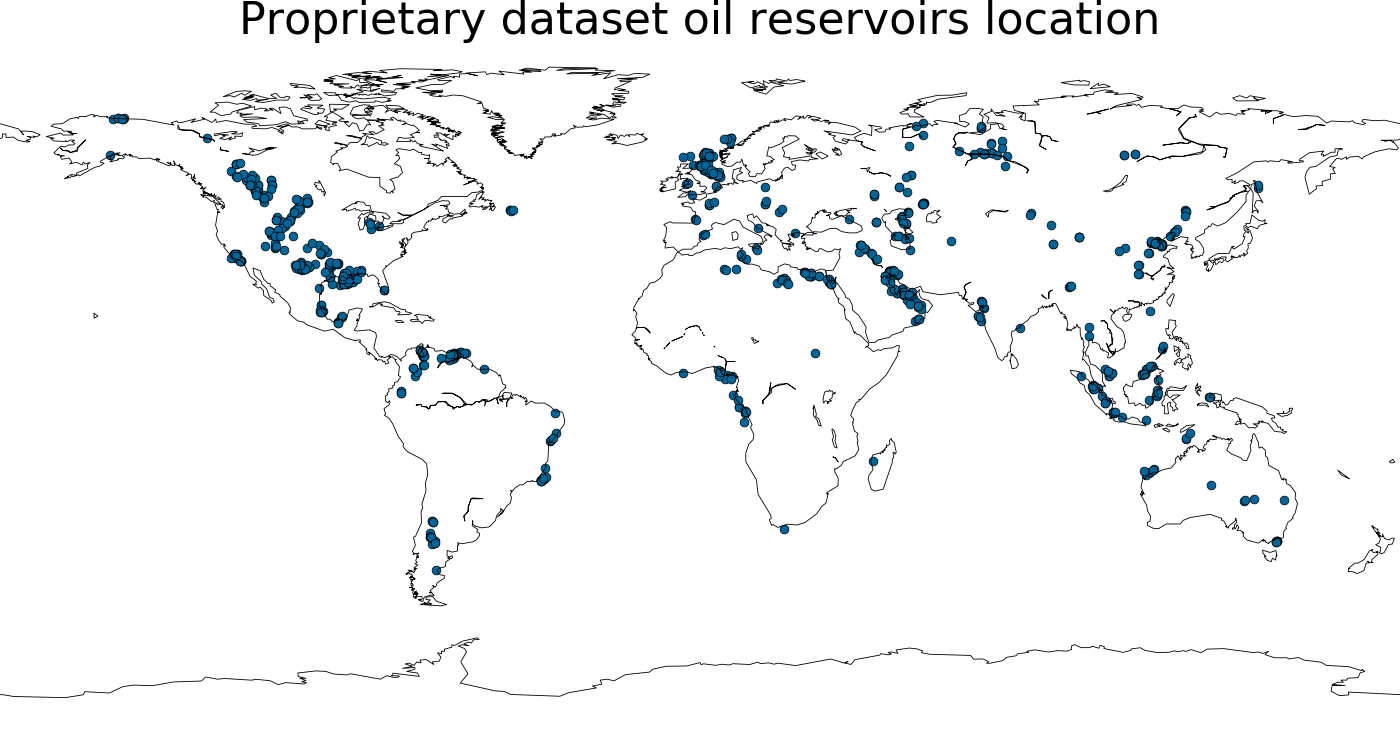}
   \caption{Oil reservoirs location from proprietary database}
   \label{fig:IP_map}
\end{figure}

Another dataset was provided by Russian oil company. It contains information about 1119 oil reservoirs throughout the world. This dataset provides more comprehensive descriptions in comparison to TORIS. Number of parameters is 199. Among  them 74 categorical and  125  numerical.  Data contains incomplete records, 38.5\% values are missing. Geographical layout is shown in \ref{fig:IP_map}.

\FloatBarrier

Only 737 of 1119 reservoirs contain expected ultimate oil recovery factor and can be used for training and testing. We group all parameters in the following way:

\begin{itemize}
\item Geometry | Seal thickness, Elevation, Water depth, True Vertical Depth, Structural dip, Closure area, Closure height, Area (original productive), Fluid contact (original OWC/GOC/GWC), Hydrocarbon column height (original oil/gas/total), Thickness (gross pay, avg/min/max), Thickness (net pay, avg/min/max)
\item Geology | Tectonic regime, Source rock depositional environment, Kerogen type, Seal rock (unit/period/epoch/age/depositional system/depositional environment lithology/classification), Structural setting, Trapping mechanism (main/secondary/tertiary), Structural compartment count, Reservoir (period/epoch/age), Depositional system (main/secondary), Depositional environment (main/secondary/tertiary),  Stratigraphic compartment count, Fracture origin (main/secondary), Lithology (main/secondary), Grain size for clastics (main/secondary), Depositional texture for carbonates (main/secondary), Depositional component for clastics/carbonates (main/secondary), Basin type, Diagenetic reservoir, Fracture reservoir type, Source rock (unit/period/epoch/age/lithology).
\item Transport, Storage and Fluid properties | API gravity,  Viscosity, Viscosity temperature, Gas specific gravity, Sulphur (\%), Wax (\%), Carbon dioxide (\%), Hydrogen sulphide (\%), Nitrogen (\%), Gas-oil ratio, FVF, Temperature, Temperature depth, Water salinity, Porosity type (main/secondary/tertiary), Porosity (matrix/fracture, avg/min/max), Permeability (air, avg/min/max), Permeability (production-derived, avg/min/max)
\item Saturations, Ratios and Pressures | Pressure (original), Pressure depth, Pressure (current), Pressure year (current), Pressure gradient (original),  Pressure (saturation), Net-gross ratio (avg), Water saturation measurement source, Water saturation (avg/min/max, \%)
\item Development and Production | Reserves (original/recoverable in-place oil/gas/condensate), Production year (cumulative), Production (cumulative oil/gas/condensate), Production year (start), Production year (plateau),  Production rate (plateau oil/gas/condensate), Production year (current), Well count (current, producing/injection), Water-cut (current, \%),  Production rate (current, oil/gas/condensate), Single well rate (max, oil/gas), Drive mechanism (main/secondary), Hydrocarbon type (main/secondary/tertiary), Improved recovery method (main/secondary/tertiary), Recovery factor (oil, primary/secondary/tertiary, \%),  Well spacing (oil/gas, average), Discovery year, Reservoir status (current),  Well count (total, production/injection), Seismic anomaly, Unconventional reservoir type
\item Location | Field name, Operator company, Reservoir unit, Basin name, Latitude, Longitude, Onshore or offshore, Country, State, Region
\end{itemize}

\section{Background}
\label{sec:background}

\subsection{Prediction interval}
The general problem statement of this study is to infer the statistical relationship between secondary recovery factor (limited to water flooding) of an oil reservoir and its available parameters. In more formal way, let $X = \{x_i\}^{n}_{i=1} \in \textbf{X}^{n} \subset \mathbb{R}^{n \times d}$ denotes numerical description of the oil reservoirs, where $n$ is the number of reservoirs and $d$ is the number of observed variables. Let $Y = \{y_i\}^{n}_{i=1} \in \mathbf{Y}^{n} \subset \mathbb{R}^{n}$ denotes column of target values. In our case $Y$ becomes the column of oil recovery factors for the corresponding reservoirs. Having available data $z_1 = (x_1, y_1), ..., z_n = (x_n, y_n)$, confidence level $\alpha$ and new oil reservoir description $x$, our purpose is to construct prediction interval $\Gamma^\alpha(z_1,...,z_n, x)$ with the following proprieties:

{\parindent0pt 
\textbf{Validity}
}

Prediction interval $\Gamma^\alpha(z_1,...,z_n, x)$ is valid if it contains true $y$ with probability not less than $\alpha$  \cite{VovkConformal2014}.

{\parindent0pt
\textbf{Efficiency}
}

Prediction interval $\Gamma^\alpha(z_1,...,z_n, x)$ is efficient if its length is relatively small and therefore informative \cite{VovkConformal2014}.

Predictive intervals are often constructed using regression models. In this study, we use decision tree based regression models.

\subsection{Regression models based on decision trees}
We apply two ensemble regression models, which use decision tree as base estimator. Since data is noisy and significant portion of values are missing it is quite natural to consider machine learning algorithms based on decision trees such as Random Forests and Gradient Boosting over decision trees. These models proved itself to be robust to noise, able to handle missing values, immune to multicollinearity and sufficiently accurate for engineering applications \cite{roy2012robustness} \cite{gomez2017study}. In this study, we provide evaluation of two tree-based algorithms: Quantile Regression Forests and Gradient boosting over decision trees with Inductive Conformal Predictors. This two approaches allow build prediction intervals as output.

\subsubsection{Quantile Regression Forests}
Random Forests were initially introduced by \citet{breiman2001random}. It is a powerful tool for high-dimensional regression and classification problems. Classical Random Forests regression tries to give a point estimate $\hat{\mu}(x)$ of the response variable $y$, given $x$. Where $\mu(x)$ is the mean of the full conditional distribution $F(y|x)$. However, it was shown that random Random Forests provides information about the full conditional distribution of the response variable \cite{meinshausen2006quantile}. \citet{meinshausen2006quantile} showed that Random Forests allows to approximate full conditional distribution $\hat{F}(y \vert x)$. Hence, having $\hat{F}(y \vert x)$ we are able to estimate quantiles of the conditional distribution as $\hat{Q}_{q}(x)$. Therefore, prediction interval could be computed as $\Gamma^\alpha(z_1,...,z_n, x) = \left[\hat{Q}_{\frac{1-\alpha}{2}}(x), \hat{Q}_{\alpha + \frac{1-\alpha}{2}}(x) \right]$. The algorithm is shown to be consistent. It was shown that under several assumptions error of the approximation to the conditional distribution converges uniformly in probability to zero for $n \rightarrow \infty$ \cite{meinshausen2006quantile}.

\subsubsection{Inductive Conformal Predictors over Gradient Boosting}
Unlike Random Forests, Gradient Boosting base estimators are trained sequentially. Each new one compensates for the residuals of the previous ones by learning the gradient of the loss function \cite{friedman2001greedy}. Since base estimators are dependent, there are no any similar way to estimate conditional distribution of the response variable as it can be done with Random Forest.

Conformal predictors is the meta-algorithm, which can be built on top of almost any supervised machine learning algorithm \cite{vovk2005algorithmic}. It allows constructing prediction set for corresponding confidence level using any regression or classification method as underlying algorithm. Conformal predictors is defined using the concept of nonconformity measure. A nonconformity measure is a measurable function $B : Z^{*} \times Z \to \mathbb{R}$, such that $B(\zeta, z)$ does not depend on the ordering of $\zeta$.
Intuition: $B(\zeta, z)$ (the nonconformity score) shows how different $z$ from the examples in $\zeta$. Possible choice: 
\begin{equation}
B(\zeta, (x, y)) = \Delta(y, \hat{f}(x))   
\end{equation}
where $\hat{f} : \textbf{X} \to \mathbf{Y'}$ --- prediction rule (Gradient Boosting here) founded from $\zeta$  as the training set and $\Delta : \textbf{Y} \times \mathbf{Y'} \to \mathbb{R}$ --- is a measure of similarity between the target and the prediction. Conformal predictors framework allows examining a range of possible target values given x by calculating its nonconformity scores. These nonconformity scores transform into so-called p-values. Range of y for which p-values does not exceed the confidence level can be presented as prediction interval. In this study, we used Inductive Conformal Predictors, which is computationally efficient modification of the initial algorithm.

Inductive Conformal Predictors framework always provides a valid prediction set under the assumption of exchangeability, which follows from i.i.d. assumption. Nevertheless, efficiency depends on chosen nonconformity measure and should be checked in each case. In our study, we use Gradient Boosting over Decision Trees as underlying algorithm. As nonconformity measure $B(\zeta, (x, y)) = \lvert y-\hat{f}(x) \rvert$ was chosen \cite{VovkConformal2014,ConformalKRR2016}.

\subsection{Clustering and visualization}

In this section we give a brief methods description, which were used in this study to identify groups of reservoirs with similar characteristics. For each group can be done a separate analysis of the oil recovery factor dependency on input parameters. Therefore, there can be identified groups with common properties for which dependency is lower or higher relative to the rest. Dimensionality reduction techniques, such as t-SNE, help to visualize high-dimensional data points preserving its spatial structure.

\subsubsection{K-means clustering}

Clustering is the technique of grouping a set of objects in high-dimensional space by distance. One of the first clustering algorithms K-means had first proposed over 50 years ago. It is still one of the most widely used algorithms for clustering \citep{hartigan1979algorithm, jain2010data}. Given $n$ data points $X = \{x_i\}^{n}_{i=1} \in \textbf{X}^{n} \subset \mathbb{R}^{n \times d}$, K-means is to group them into $k$ clusters.
At initial step random $k$ data points are selected as cluster means. Then the two following steps are repeated until convergence. The first step is to assign each data point to cluster with nearest mean. The second step is to recalculate cluster means corresponding to the new partition. In our work we used an extension of K-means: K-means++. It specifies a procedure to initialize cluster means \cite{arthur2007k}. Proposed initializing procedure makes algorithm stable and provides an optimal solution with strong chance.

\subsubsection{t-SNE}

One of the common ways to visualize high-dimensional data is to find transformation from initial space into two or three-dimensional space with preserving spatial relationships. Well known PCA technique provides linear transformation to low dimensional space by finding the projections, which maximize the variance. However, there are more effective non-linear methods to visualize the spatial structure \citep{li2017application}. T-distributed Stochastic Neighbor Embedding (t-SNE) is a non-linear dimensionality reduction method, which is widely used for high-dimensional datasets visualizing \cite{maaten2008visualizing}. The t-SNE algorithm has two main steps. First, t-SNE generates a probability distribution over pairs of high-dimensional objects in such a way that similar objects will be selected with a high probability, while the probability of selecting dissimilar points will be low. The t-SNE then determines a similar probability distribution over points in low-dimensional space and minimizes the Kullback-Leibler divergence between the two distributions, taking into account the position of the points.
In other words, algorithm finds a map, which placed similar objects nearby in a low-dimensional space while keeping dissimilar objects well separated. Thus, this method is suitable to visualize clustering structure.

\section{Methodology}
\label{sec:methodology}

 Usually, reserves-estimation methods are divided into two classes: for pre-production phases and post-production phases. The main difference between these two classes is the type of input data. Methods related to pre-production phases usually predate development planning \citep{demirmen2007reserves}. These methods generally entail more significant errors and uncertainty. The economic effect can be greater compared to post-production techniques. Similarly, we build and evaluate two data-driven models. For both models we consider recovery factor limited to water flooding (secondary method). First one takes a set of parameters available at pre-production steps as input. The second one takes an extended set of parameters as input, including production data and development scheme description. We evaluate models with commonly used cross-validation technique. In the following sections, we describe both models design and evaluation details.

\subsection{Model for pre-production phases}

At pre-production phases, one of the main objective is to estimate economic potential of the oilfield. Expected ultimate oil recovery factor estimation is an essential step for asset valuation. The pre-production model is supposed to be used during reservoir exploration when often all available information is just averaged reservoir characteristics, which can be estimated by measuring the characteristics at several appraisal wells, as well as using seismic exploration. Such parameters may refer to reservoir geometry, geology, transport and fluid properties. Using this data reserves need to be assessed as accurately as possible. However, recovery factor strongly depends on economic effect that is difficult to forecast. Hence, training set should contain oil reservoirs, which were in development with diverse economic environments and with different technologies. In order to increase training set diversity, we build a training set using two data sources: TORIS dataset and proprietary dataset. We identified similar parameters of these two sources and converted measurements to common units. We divide parameters in the combined dataset into two groups:

\begin{itemize}
\item Not suitable for pre-production model input | Production rate (current, oil tons per day), Well spacing (field averaged, $km^2$), Pressure (current, atm), Well count (total production), Well count (total injection), Production (cumulative oil, mln tons)
\item Suitable for pre-production model input | Thickness (net pay average, m), Net/gross ratio (average), Porosity (matrix average, \%), Water saturation (average, \%), FVF (oil, $m^3$ in reservoir conditions/$m^3$ in standard conditions), Depth (top reservoir, m TVD), Temperature (original, deg. C), Pressure (original, atm), Permeability (air average, mD), Reservoir age (mln years), API gravity (average, tons/$m^3$), Viscosity (cp), Water salinity (ppm), Reserves (original oil in-place, mln tons), Gas/oil ratio (initial, $m^3$ in standard conditions/tons), Lithology (main), Structural dip (degree)
\end{itemize}

Parameters from the first group usually are not available at pre-production phases, while parameters from the second group could be estimated with several appraisal wells.

Both sources contain estimated expected ultimate (primary+secondary) oil recovery factors. The more depleted reservoir, the more reliable recovery factor estimation we have in the dataset. The purpose is to develop technique for estimating actual economic potential of the oil reservoir. Therefore, we consider only reservoirs, which are close to depletion ($\geq$90\% of oil reserves have been extracted). Preliminary experiments show that removing lines with more than one missing value and leaving only reliable RFs, we did not lose the accuracy of the model on cross-validation. On the other hand, we got the opportunity to study the structure of the data in a multidimensional space. But we do not get rid of all the gaps in the data, so we use the missing values handling mechanism. Preliminary experiments have shown that the decision tree missing values handling mechanism coped best with missing values. It works as follows: the tree decides whether missing values go into the right or left node. It chooses which to minimise loss. This approach treats missing values as missing at reason or missing not at random. It allows capturing the signal in the missing data distribution \cite{twala2008good}.

The next step is to build and evaluate Quantile Regression Forests and Gradient Boosting with ICP using filtered dataset. We calculate error metrics on cross-validation and check validity and efficiency. Also, we perform feature importance analysis using best model. We consider F-score within the tree-based model. F-score is widely used in similar applications for feature importance analysis \cite{orlov2019advanced, erofeev2019prediction}. F-score represents the number of times a feature is used to split the data across all trees in the ensemble.

To perform cluster analysis and to analyze the spatial structure of the data, we consider records with not more than one missing value. Cluster analysis helps to find groups of objects (clusters) that are similar to each other. As a similarity measure, we use euclidean distance for scaled parameters. We perform cluster analysis and identify the number of clusters. We visualize spatial structure with t-SNE algorithm and provide analysis of parameters distributions for each cluster. We evaluate metrics of tree-based models for each cluster separately and compare results. Finally, we compare the parameters distributions within clusters and make a conclusion.

\subsection{Model for post-production phases}

During post-production phases, additional information about an oil reservoir becomes available, such as development details, production dynamic and other measurements. We consider the proprietary database as the only source to form training set. It includes a more detailed oil reservoirs descriptions with timestamps of the measurements made during the production phases. We consider items with more than 50\% missing values as non-informative. These items negatively affect the quality of the model if we include them in the training set. The following parameters have selected as input:

Onshore or offshore, Elevation (m),  Water depth (m), Hydrocarbon type (main), Discovery year, Reservoir status, Well count (total production), Well count (total injection), Seismic anomaly, True vertical depth (top of reservoir, m), Structural dip (degree), Area (original productive, $km^2$), Fluid contact (original, m TVD), Hydrocarbon column height (original oil, m), Hydrocarbon column height (original total, m), Reservoir age, Depositional system (main), Depositional environment (main), Stratigraphic compartment count, Lithology (main), Thickness (gross average, m), Net/gross ratio (average), Thickness (net pay average, m), Porosity (matrix avg, min, max \%), Permeability (air avg, min, max mD), Water saturation (avg, min, max \%), TOC (avg, min, max \%), Kerogen type, Reserves (original in-place oil, mln tons), Reserves (original in-place oil equivalent, mln tons), Production year (cumulative), Production (cumulative oil, mln tons), Production year (start), Well count (current producing), Well count (current injection), Water-cut (current \%), Production rate (current oil, tons per day), API gravity (average deg. API), Viscosity (cp), Viscosity temperature (deg. C), Sulphur (\%), Gas/oil ratio (initial m3 in standard conditions / tons), FVF (oil m3 in reservoir conditions / m3 in standard conditions), Temperature (original deg. C), Temperature depth (m), Pressure (original atm), Pressure depth (m), Pressure (current atm), Pressure gradient (original atm/m), Pressure (saturation atm), Water salinity (ppm), Improved recovery method (secondary), Well spacing (oil average $km^2$)

\begin{figure}[ht!]
  \centering
    \includegraphics[width=.6\textwidth]{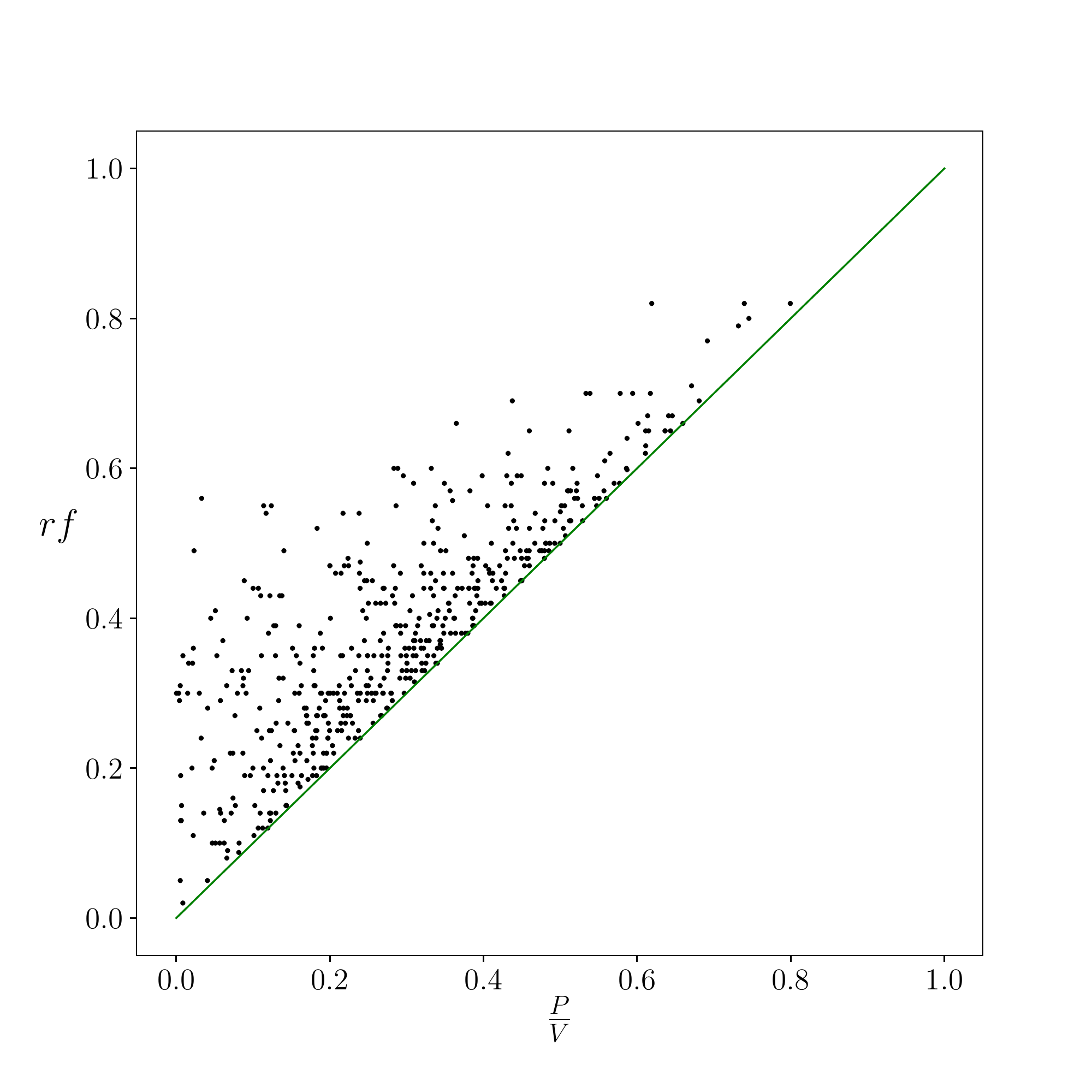}
    \caption{Scatter plot shows that ratio of cumulative oil production ($P$) and original oil in place ($V$) gives a close lower bound to oil recovery factor ($rf$).}
    \label{fig:rf_vs_pv}
\end{figure}

We selected parameters that affect the oil recovery factor, at least indirectly. Some of the parameters could be correlated. Multicollinearity particularly not affect random forest or tree-based gradient boosting algorithms by nature. On the contrary, the algorithm learns to use a secondary correlated feature if there is a missing primary one \cite{kotsiantis2013decision}. We use target encoding technique to transform categorical parameters in a numeric form. It replaces each categorical parameter value by average oil recovery factor among this category. Similarly to pre-production model case we use decision tree missing values handling mechanism. A ratio of cumulative oil production ($P$) to original oil in place ($V$) gives a lower bound to estimated oil recovery factor ($rf$) as shown in Figure \ref{fig:rf_vs_pv}. General analysis of the production behaviour of all oil reservoirs in training set can help to estimate margin between $\frac{P}{V}$ and $rf$. Denote $\Delta t$ as difference between ``Production year (cumulative)'' and ``Production year (start)''. Then, approximation of the function $\hat{f}(\Delta t, V, \textbf{w}^{\ast}) \approx \frac{P(\Delta t)}{rf*V}$ could give an approximation of the oil recovery factor as $\hat{rf}(P(\Delta t), \Delta t, V) \approx \frac{P(\Delta t)}{\hat{f}(\Delta t, V, \textbf{w}^{\ast})*V}$, where \textbf{w} are tunable parameters, which can be selected with mean squared error minimization

\begin{equation}
\mathbf{w}^{\ast} = \underset{\mathbf{w}}{\mathrm{arg}\min}\frac{1}{n}\sum_{i=1}^{n}(\frac{P_{i}(\Delta t_{i})}{V_{i}*\hat{f}(\Delta t_{i}, V_{i}, \textbf{w})} - rf_i)^2.    
\end{equation}

$\hat{f}(\Delta t, V, \textbf{w}^{\ast})$ should meet the same conditions as $\frac{P(\Delta t)}{rf*V}$, i.e.

\begin{align}
&\hat{f}(\Delta t, V, \textbf{w}^{\ast}) \rightarrow 1,\ \ \Delta t \rightarrow + \infty, \\
&\hat{f}(0, V, \textbf{w}^{\ast}) = 0.
\end{align}

This approach allows us to generate informative extra features for any machine learning algorithm using different functional families. Similarly to known cumulative oil production curves equations \cite{fetkovich1996useful}, we consider $\hat{f}(\Delta t, V, \textbf{w}^{\ast})$ as exponential and hyperbolic functional families. Tunable parameters $\textbf{w}^{\ast}$ can be found by minimizing error on a training set, then extra feature can be calculated both for training and test set as $\hat{rf}(P(\Delta t), \Delta t, V)$.

We combine the approximation curves approaches with tree-based ensembles using stacking. Stacking is an efficient method for combining several machine learning algorithm in which the outputs, generated by several first-level models are used as inputs for second-level model. We use approximations obtained from general production curves as input for tree-based ensemble. We also perform feature importance analysis using best model. We consider F-score within the tree-based model similarly to pre-production model case.

Overall, the first step is to make data preprocessing in the way described above. The second step is to conduct an analysis of the production dynamic and select appropriate functional families for generating extra features to improve predictive models accuracy. The third step is evaluation and comparison of the Quantile Regression Forest and Gradient Boosting over decision trees with Inductive Conformal Predictors using different extra features subsets.

\subsection{Evaluation metrics}

To evaluate accuracy, validity and efficiency of the considering models we use cross-validation. Cross-validation is primarily used in machine learning to estimate generalization ability of the algorithm. The procedure is as follows: randomly shuffle the data, split the data into $K$ groups, each group is considered as a test set and the remaining part --- as a train set. Since each data point would be considered as a test point, we obtain predictions of a model for all datapoints. Denote $\hat{y}$, $\hat{l}^{\alpha}$ and $\hat{u}^{\alpha}$ as vectors of predictions, lower bounds and upper bounds of the prediction intervals on confidence level $\alpha$ obtained on cross-validation respectively. We use two regression metrics $R^2$ (Coefficient Of Determination) and MAE (Mean Absolute Error).$R^2$ is a dimensionless value that shows how much better the algorithm predicts than the trivial prediction with the sample mean as the prediction (1 - perfect prediction, 0 - the algorithm predicts in average with the same squared error as the mean prediction, and $<$ 0 - the model has a greater error than the mean prediction). MAE - mean absolute error, has the same dimension as the target variable. This two metrics calculated on cross-validation give an objective assessment of the model. Metrics can be expressed in the following form:

\begin{align}
&Mean\ average\ error\ (MAE): MAE(\hat{y}, y) = \frac{1}{n}\sum_{i=1}^{n}\lvert y_i - \hat{y}_i \rvert, \\
&Coef.\ of\ determination\ (R^2): R^2(\hat{y} ,y) = 1 - \frac{\sum_{i=1}^{n}(\hat{y}_i - y_i)^2}{\sum_{j=1}^{n}(\frac{1}{n}\sum_{k=1}^{n} y_k - y_j)^2}.
\end{align}

To evaluate the validity of the prediction intervals we calculate the coverage rate, which should be greater or equal than confidence level $\alpha$:

\begin{equation}
\frac{1}{n} \sum_{i=1}^{n} \mathbb{I}_{\hat{l}_{i}^{\alpha} \leq y_i \leq \hat{u}_{i}^{\alpha}}.
\end{equation}

To evaluate the efficiency of the prediction intervals, we calculate its mean width. The less mean width, the more informative prediction intervals.

\begin{equation}
\frac{1}{n}\sum_{i=1}^{n}\lvert \hat{u}_{i}^{\alpha} - \hat{l}_{i}^{\alpha} \rvert
\end{equation}

\section{Results}
\label{sec:results}
\subsection{Model for pre-production phases}

\begin{figure}[!h]
\begin{minipage}{.5\linewidth}
\centering
\subfloat[]{\label{clustering:a}\includegraphics[scale=.3]{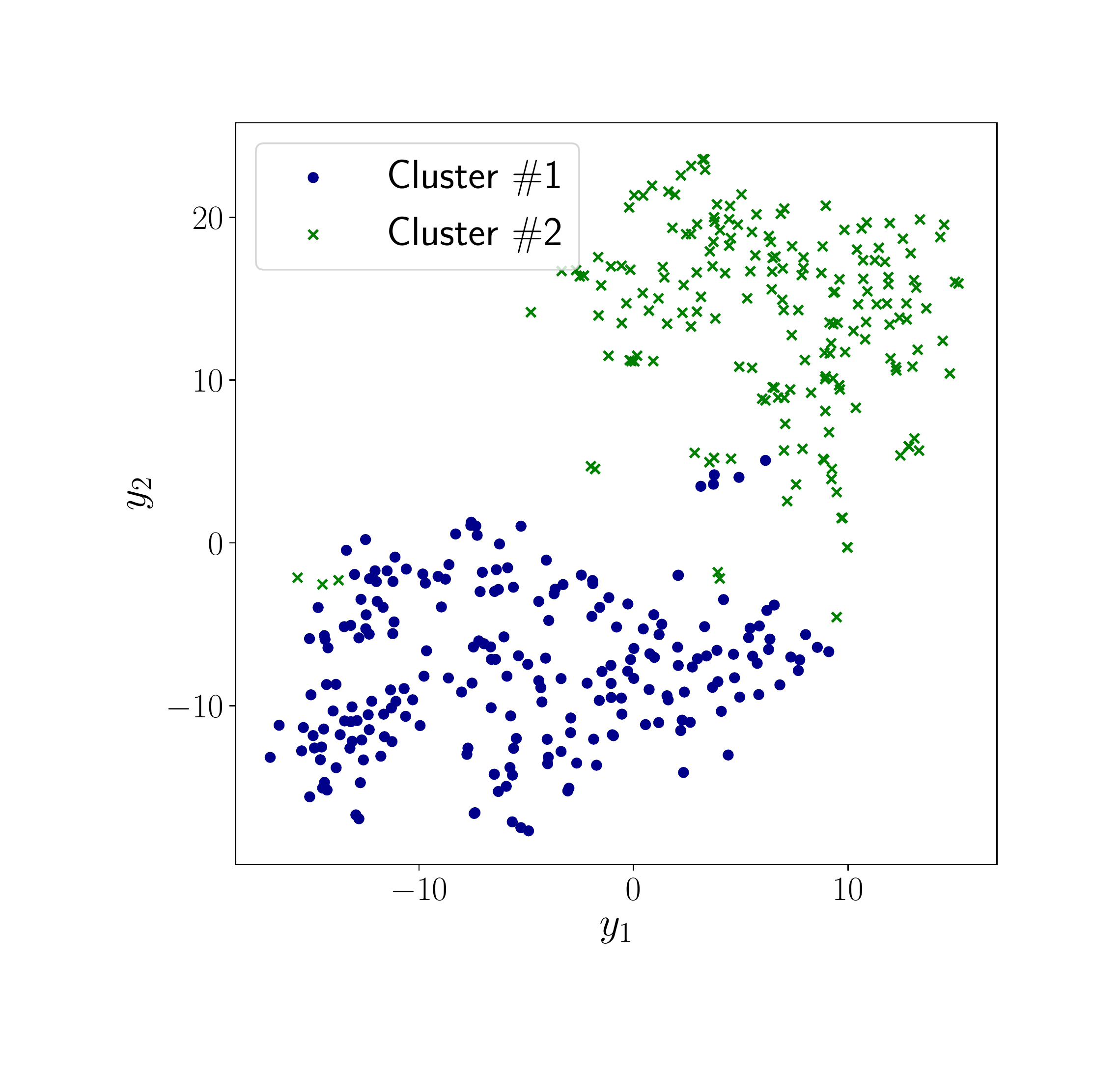}}
\end{minipage}
\begin{minipage}{.5\linewidth}
\centering
\subfloat[]{\label{clustering:b}\includegraphics[scale=.3]{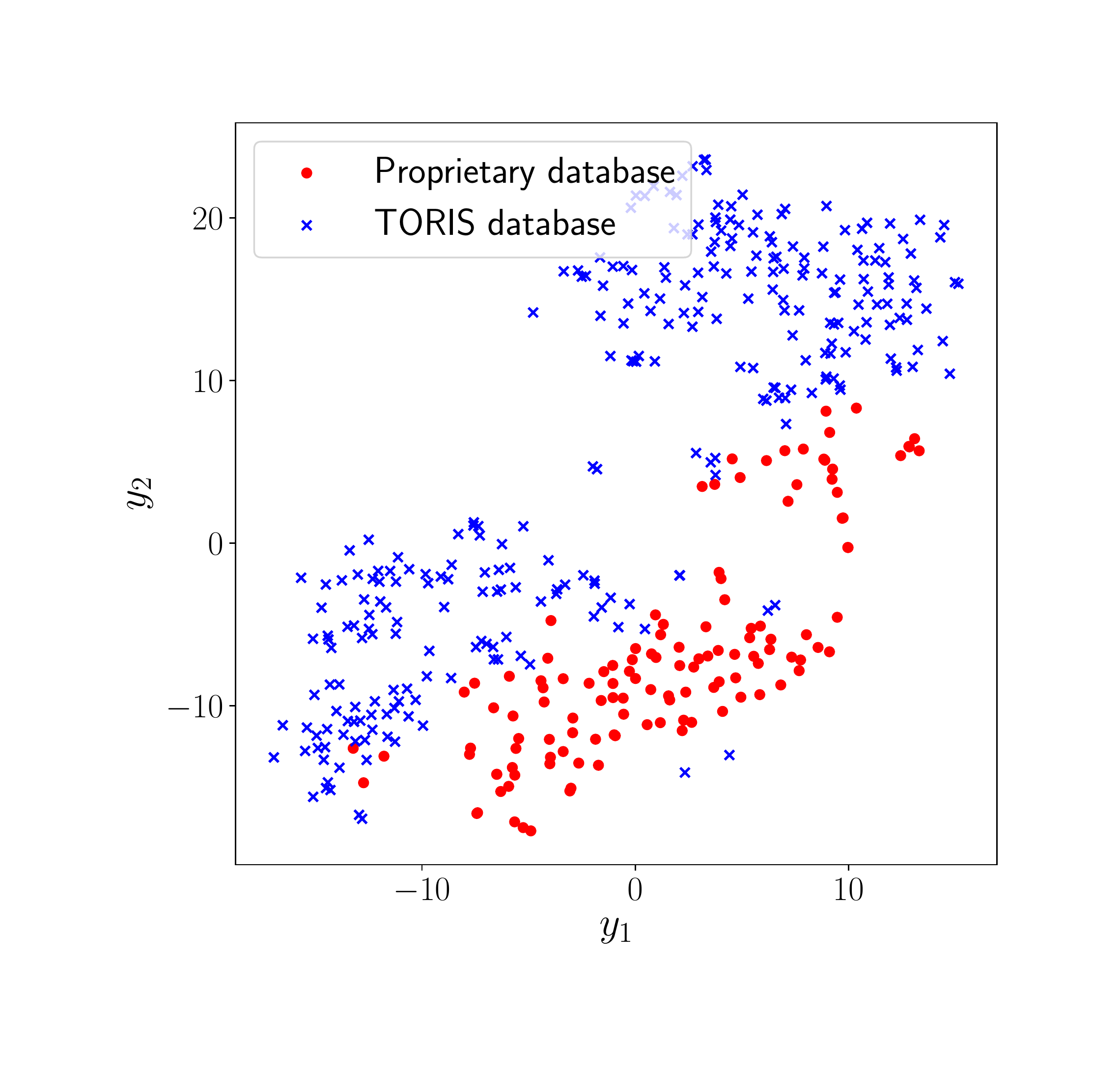}}
\end{minipage}\par\medskip
\centering
\subfloat[]{\label{clustering:c}\includegraphics[scale=.3]{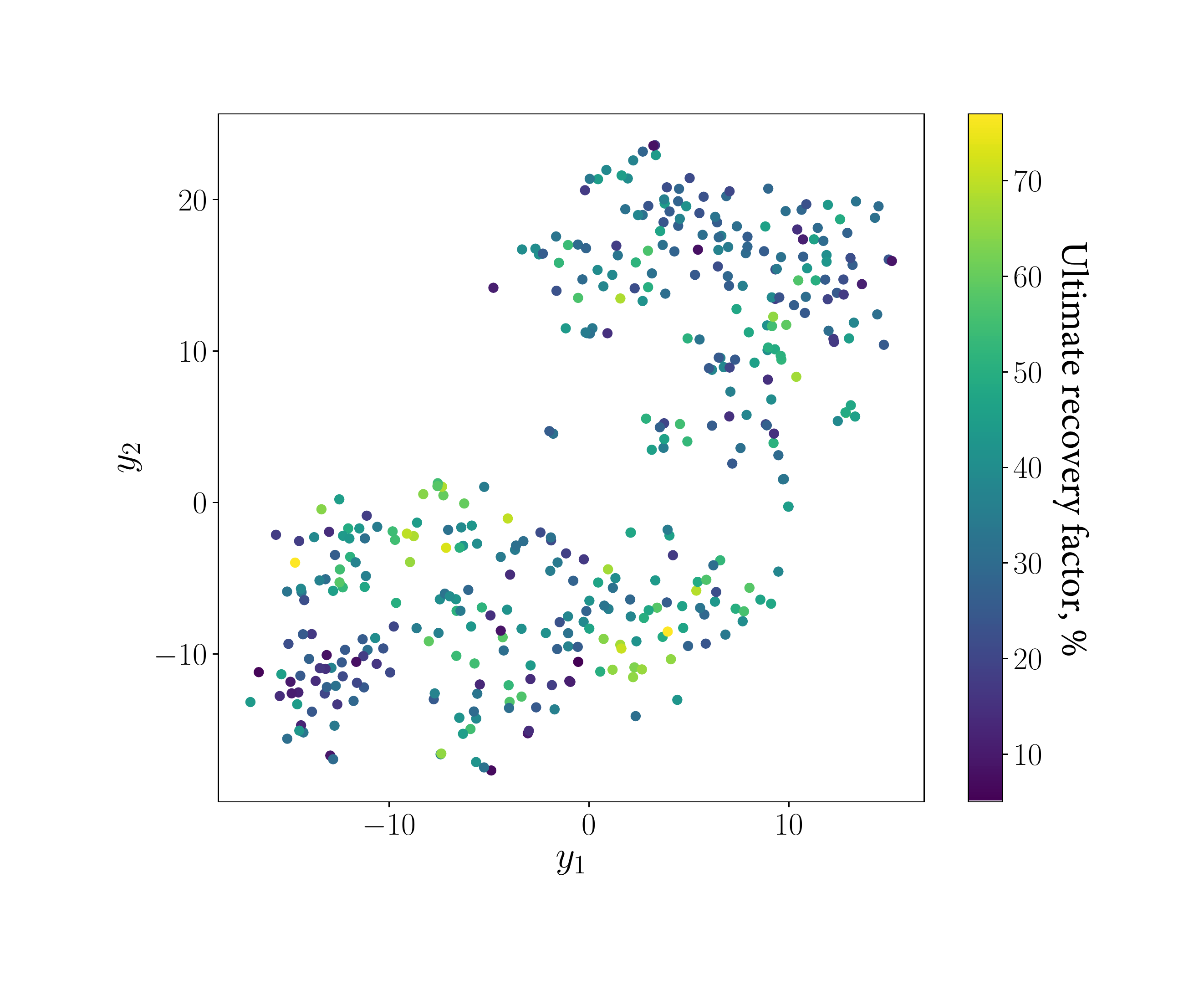}}

\caption{Two dimensional t-SNE training set embeddings visualisations. T-sne transforms training set into two-dimensional space with preserving spatial relationships. Figure \ref{clustering:a} demonstrates result of K-means partitioning in original space. K-means partition is consistent with the observed cluster structure in embedded space. Figure \ref{clustering:b} demonstrates data sources of the oil reservoirs descriptions. Figure \ref{clustering:c} demonstrates distribution of expected ultimate oil recovery factor.}
\label{fig:clustering}
\end{figure}

\begin{figure}[!h]
  \centering
    \includegraphics[width=.9\textwidth]{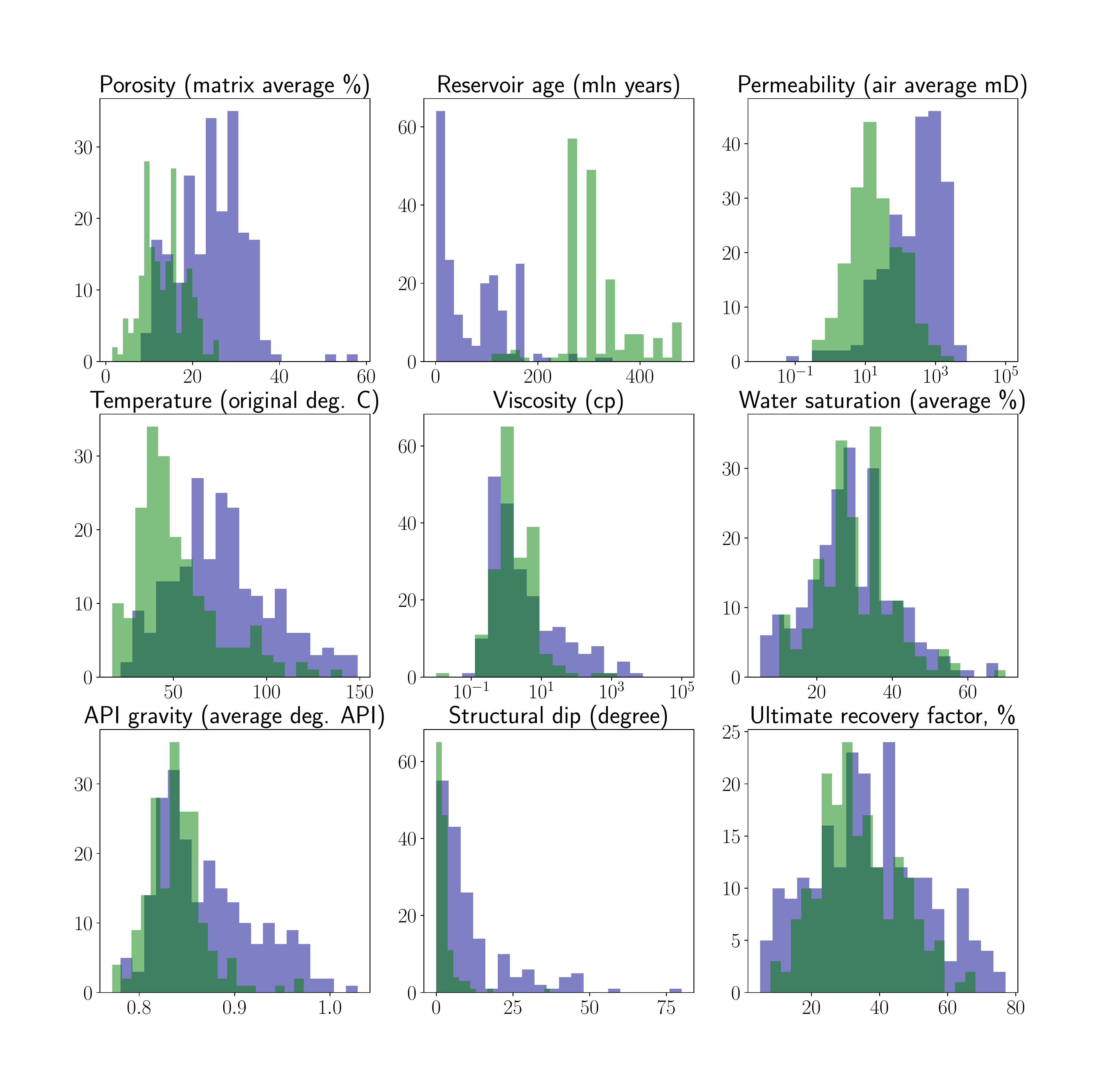}
    \caption{Histograms demonstrate most distinguishable parameters distributions for each cluster. Blue color - histograms for cluster \#1. Green color - histograms for cluster \#2.}
    \label{fig:clustering_distributions}
\end{figure}

\begin{figure}[!h]
  \centering
    \includegraphics[width=1.\textwidth]{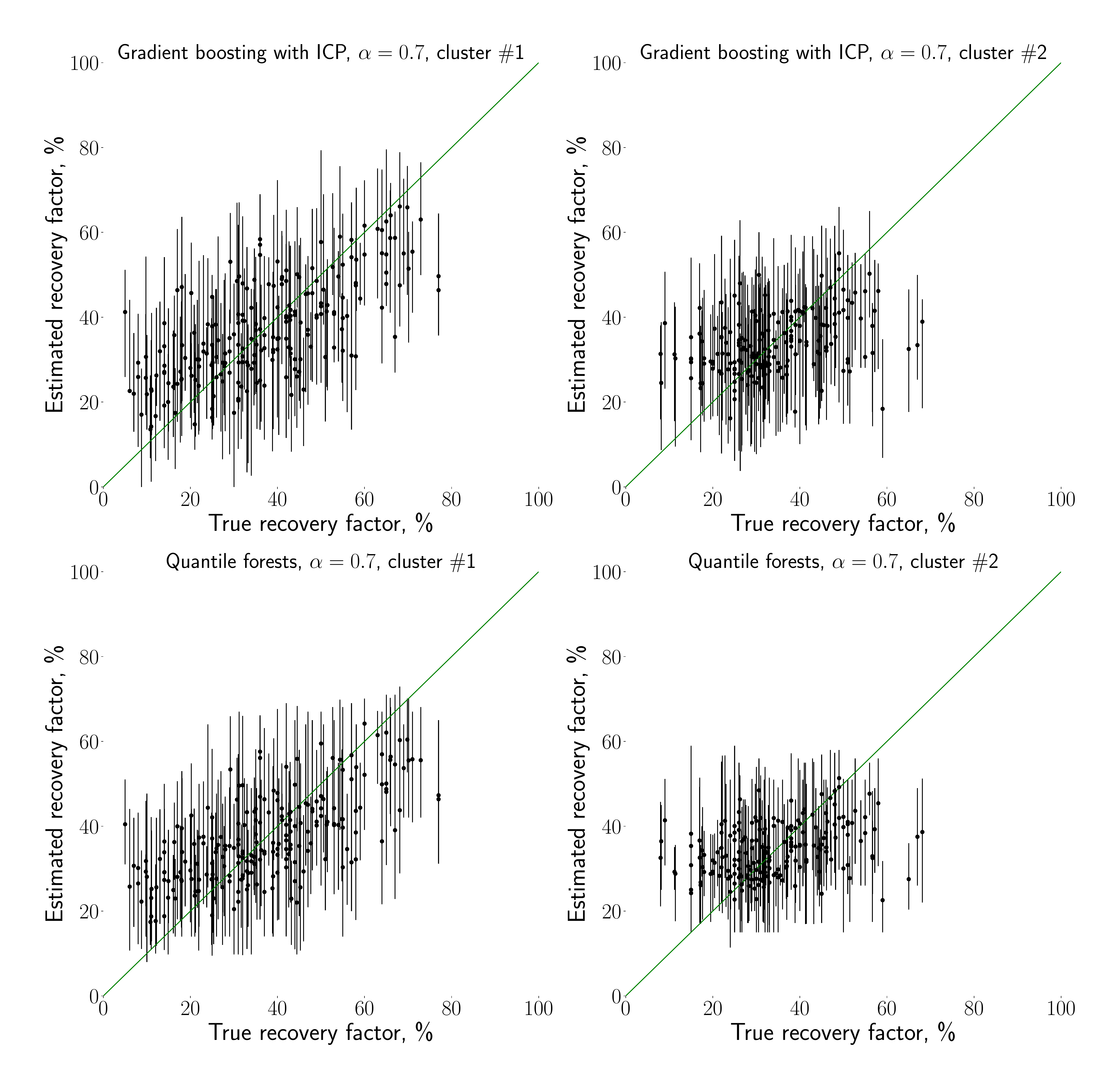}
    \caption{Prediction intervals visualization for 70\% confidence level. The graphs show randomly picked 25\% points from leave one out cross-validation.}
    \label{fig:time_independent_confidence}
\end{figure}

To build the prediction model for pre-production phases, we combine and filter both datasets: TORIS and proprietary. The resulting training set contains 407 oil reservoirs, described by 16 time-independent parameters. Table \ref{table:ep_cv_results} demonstrates relatively low pre-production phase model's accuracy for the whole dataset. This fact led us to analyze if there are subset(s) presence for which dependence between parameters and recovery factor stronger than for others. However, to iterate over all subsets is the problem with exponential computational complexity. Thus, we decided to analyze the data's cluster structure in high dimensional space, where each reservoir represented as a numerical vector of its parameters. We perform an analysis of the spatial structure of the training set using clustering technique K-means and t-SNE algorithm for data visualization. Cluster analysis indicated the presence of two clusters in original space. Two-dimensional training set point embeddings from the training set obtained with t-SNE depicted in Figure \ref{fig:clustering}. This embeddings visualization confirms the presence of two clusters. K-means partition is consistent with the observed cluster structure in embedded space, Figure \ref{clustering:a}. Figure \ref{clustering:b} shows that cluster \#1 contains oil reservoirs from Proprietary dataset as well as from TORIS dataset. Cluster \#2 contains oil reservoirs primarily from TOIRS. Therefore, reservoirs from cluster \#2 geographically located primarily in North America. On the other hand, reservoirs from the cluster \#1 have more geographic diversity. Comparison of parameters distributions for each cluster shows that cluster \#1 contains reservoirs with higher porosity and permeability, than cluster \#2. A significant difference is observed in the geological age of the rock (see Figure \ref{fig:clustering_distributions}). Also, clusters differ in main sediments. Most of the reservoirs from cluster \#1 (98 \%) are terrigenous. By contrast, most of the reservoirs from cluster \#2 (65 \%) are carbonate.  On both clusters, as well as on entire training set, we evaluate the considered machine learning algorithms. Table \ref{table:ep_cv_results} and Figure \ref{fig:time_independent_confidence} demonstrate that recovery factor predictions are more accurate for cluster \#1. For the first cluster, $R^2$ is closer to 0.5, which indicates that the dependence of the oil recovery factor on the input parameters is captured by the model for reservoirs from this cluster. For the second cluster, $R^2$ does not differ much from 0; it says that the model's predictions are not much better than the simplest mean predictions. One could spot that MAE for the first cluster is higher than for the second. This is due to the fact that the RF range for the first cluster is wider than for the second. Therefore, in this case, the predictive ability with $R^2$ is more appropriate. Based on these facts, we can assume that for terrigenous reservoirs with high porosity and permeability, the oil recovery factor's dependence on presented input parameters is much higher. Gradient Boosting with ICP and Quantile Regression Forests both provides prediction intervals close to valid, Table \ref{table:independent_intervals}. Gradient Boosting with ICP demonstrates the most accurate result, Table \ref{table:ep_cv_results}.

\FloatBarrier

\FloatBarrier
\begin{table}[!h]
\begin{center}
\begin{tabular}{|c||c|c|c|}
\hline
\multicolumn{4}{|c|}{Gradient Boosting} \\
\hline
Cluster \# & 1 & 2 & 1\&2 \\
\hline
\hline
MAE & 9.57 & 8.63 & 9.06 \\
$R^2$ & 0.47 & 0.1 & 0.38 \\
\hline
\hline
\multicolumn{4}{|c|}{Random Forests} \\
\hline
Cluster \# & 1 & 2 & 1\&2 \\
\hline
\hline
MAE & 9.93 & 8.80 & 9.13 \\
$R^2$ & 0.45 & 0.09 & 0.38 \\
\hline
\end{tabular}
\caption{Comparison of Gradient Boosting and Random Forests algorithm on both clusters, as well as on entire training set. Error metrics calculated on leave one out cross-validation. Listed metrics demonstrate more accurate results for cluster \#1 in terms of $R^2$. Mean absolute error for cluster \#2 is less than for the cluster \#1, but that is due to lower oil recovery factor range in the cluster \#2.}
\label{table:ep_cv_results}
\end{center}
\end{table}

\begin{table}[!h]
\begin{center}
\begin{tabular}{|c||c|c|c||c|c|c||c|c|c|}
\hline
\multicolumn{10}{|c|}{Gradient Boosting with ICP} \\
\hline
Cluster \# & \multicolumn{3}{c||}{1} & \multicolumn{3}{c||}{2} & \multicolumn{3}{c|}{1\&2} \\
\hline
$\alpha$ & 0.7 & 0.8 & 0.9 & 0.7 & 0.8 & 0.9 & 0.7 & 0.8 & 0.9 \\
\hline
\hline
Mean width & 24.91 & 31.36 & 42.05 & 25.80 & 33.49 & 47.03 & 24.69 & 30.41 & 40.41 \\
Coverage & 0.66 & 0.77 & 0.87 & 0.72 & 0.86 & 0.95 & 0.70 & 0.81 & 0.91 \\
\hline
\hline
\multicolumn{10}{|c|}{Quantile Regression Forests} \\
\hline
Cluster \# & \multicolumn{3}{c||}{1} & \multicolumn{3}{c||}{2} & \multicolumn{3}{c|}{1\&2} \\
\hline
$\alpha$ & 0.7 & 0.8 & 0.9 & 0.7 & 0.8 & 0.9 & 0.7 & 0.8 & 0.9 \\
\hline
\hline
Mean width & 25.75 & 31.66 & 39.85 & 20.28 & 25.13 & 32.74 & 23.50 & 29.17 & 37.10\\
Coverage & 0.70 & 0.82 & 0.87 & 0.66 & 0.77 & 0.86 & 0.71 & 0.81 & 0.89\\
\hline
\end{tabular}
\caption{Mean width and coverage rate calculated on leave one out cross-validation.}
\label{table:independent_intervals}
\end{center}
\end{table}

\begin{figure}[!h]
  \centering
    \includegraphics[width=1.\textwidth]{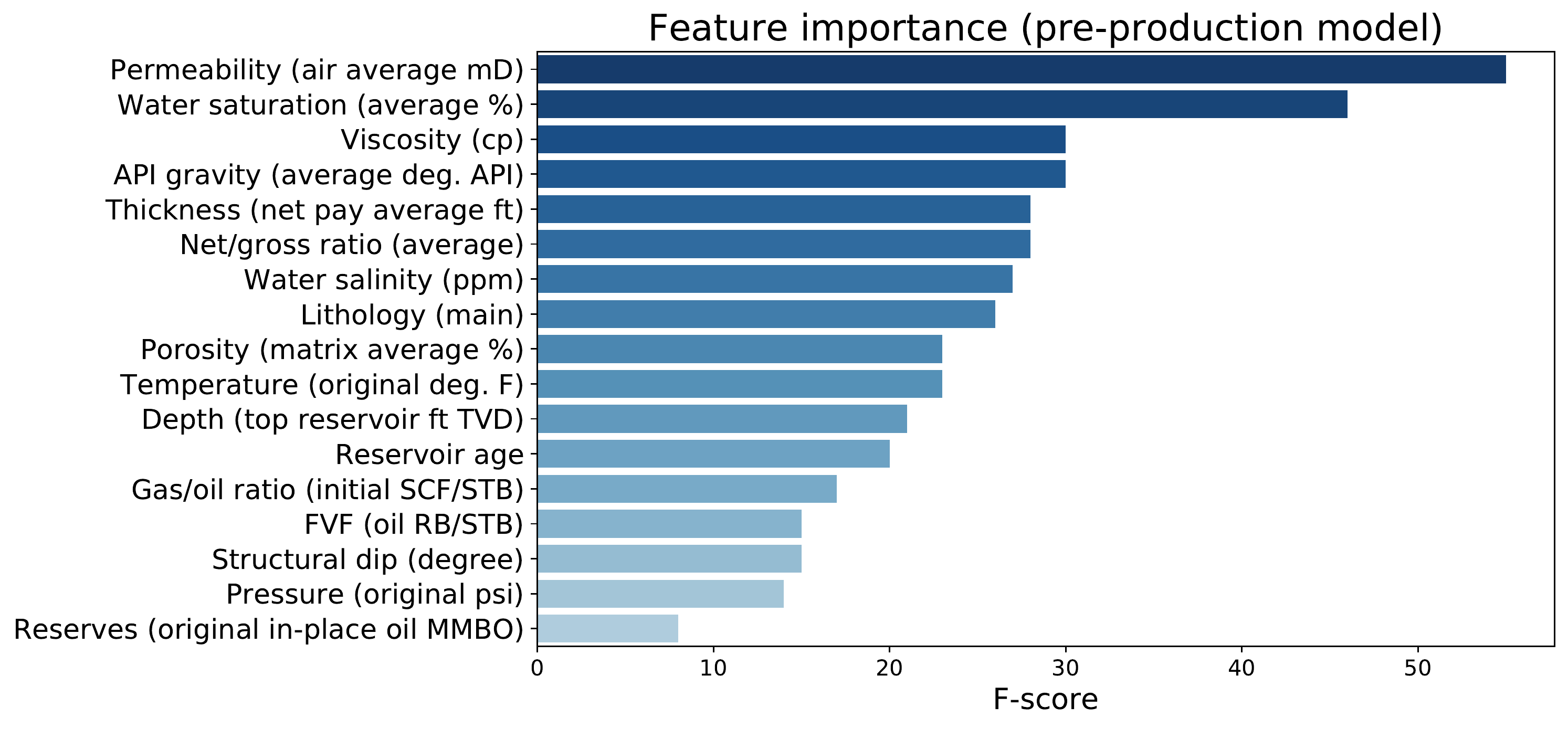}
    \caption{Feature importance for the final Gradient Boosting pre-production model. Importance represents as F-score -  the number of times a feature is used to split the data across all trees in ensemble.}
    \label{fig:fi_preprod}
\end{figure}
\FloatBarrier

\subsection{Model for post-production phases}

To build the prediction model for post-production phases, we use proprietary database only since it contains a comprehensive description of oil reservoirs as well as timestamps of the current measurements. The resulting training set contains 549 oil reservoirs, each of which is described by 67 parameters. Due to a large number of parameters (curse of dimensionality) and a high portion of the missing values, we do not perform cluster analysis. First of all, we analyzed the accuracy of recovery factor estimation with general production curves $\hat{f}(\Delta t, V, w^\ast)$ as an approximation of $\frac{P(\Delta t)}{rf*V}$. We search approximation in exponential and hyperbolic functional families. The first group of functional families depends only on $\Delta t$:

\begin{align}
&f_{hyp}(\Delta t, w) = \frac{\Delta t}{\Delta t + w},\\
&f_{exp}(\Delta t, w) = 1 - e^{-\frac{\Delta t}{w}}.
\end{align}

For both families, parameter $w$ determines the slope of the curve. Figure \ref{exp_splitV:a} demonstrates optimal curve from exponential functional family. Since reservoirs with large original oil in place are deplete more slowly, their production curve slope will be more shallow. Figure \ref{exp_splitV:b} shows that parameter $w$ directly depends on original oil in place ($V$). Figure \ref{exp_splitV:c} demonstrate type of this dependency.  Thus, we consider the second group of functional families. Adding dependence on original oil in place ($V$), we obtain more complex models:

\begin{figure}[!h]
\centering
\subfloat[]{\label{exp_splitV:a}\includegraphics[scale=.3]{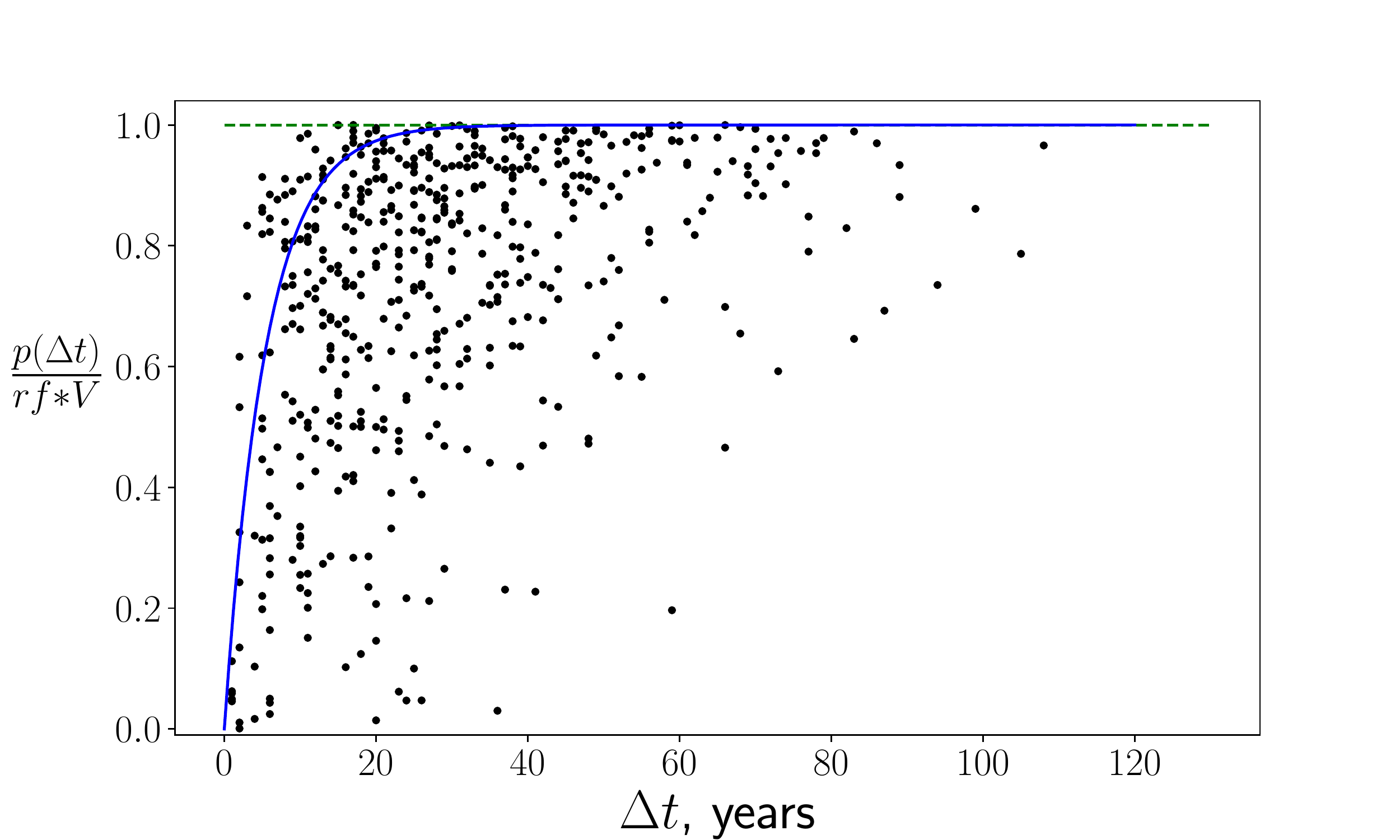}}\par\medskip
\centering
\subfloat[]{\label{exp_splitV:b}\includegraphics[scale=.3]{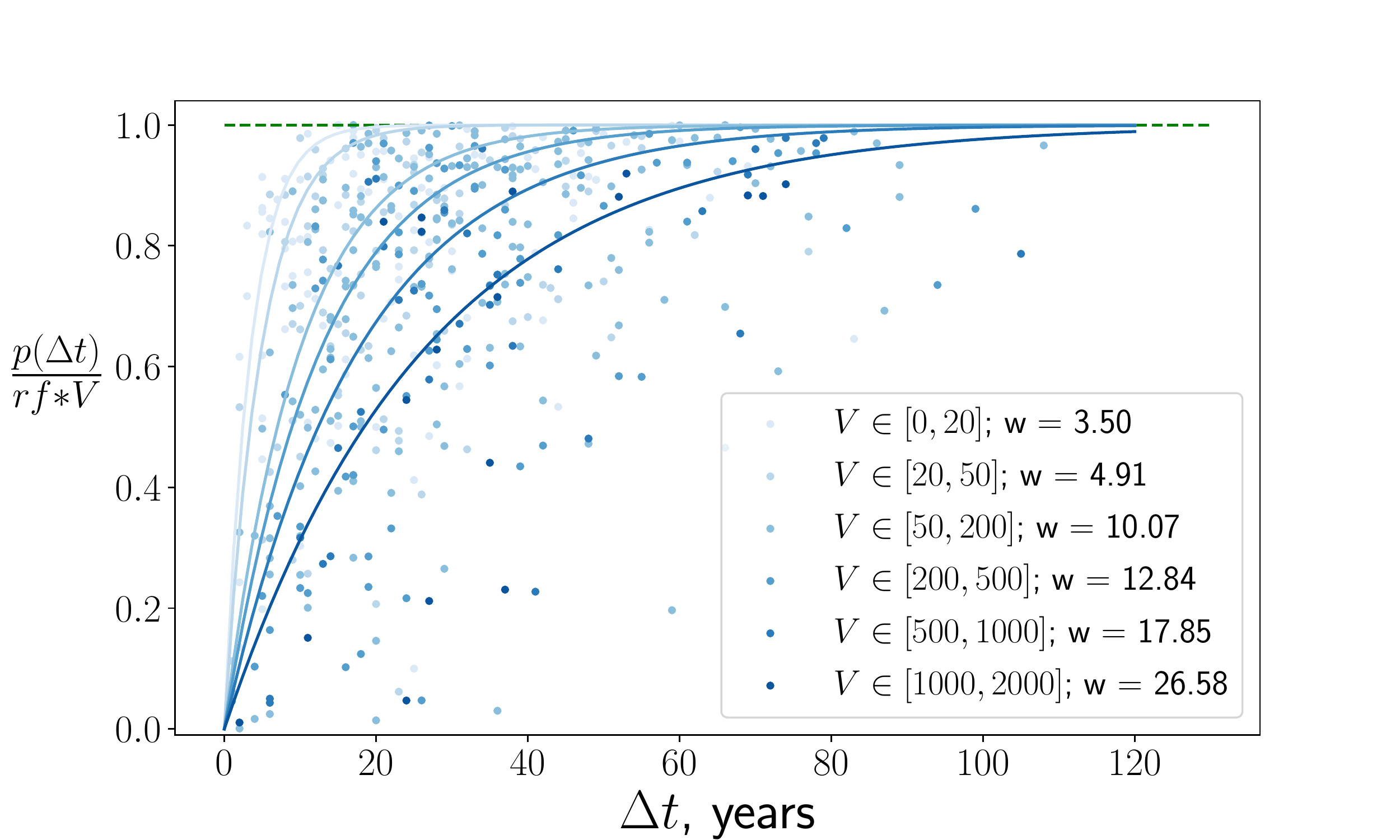}}\par\medskip
\centering
\subfloat[]{\label{exp_splitV:c}\includegraphics[scale=.3]{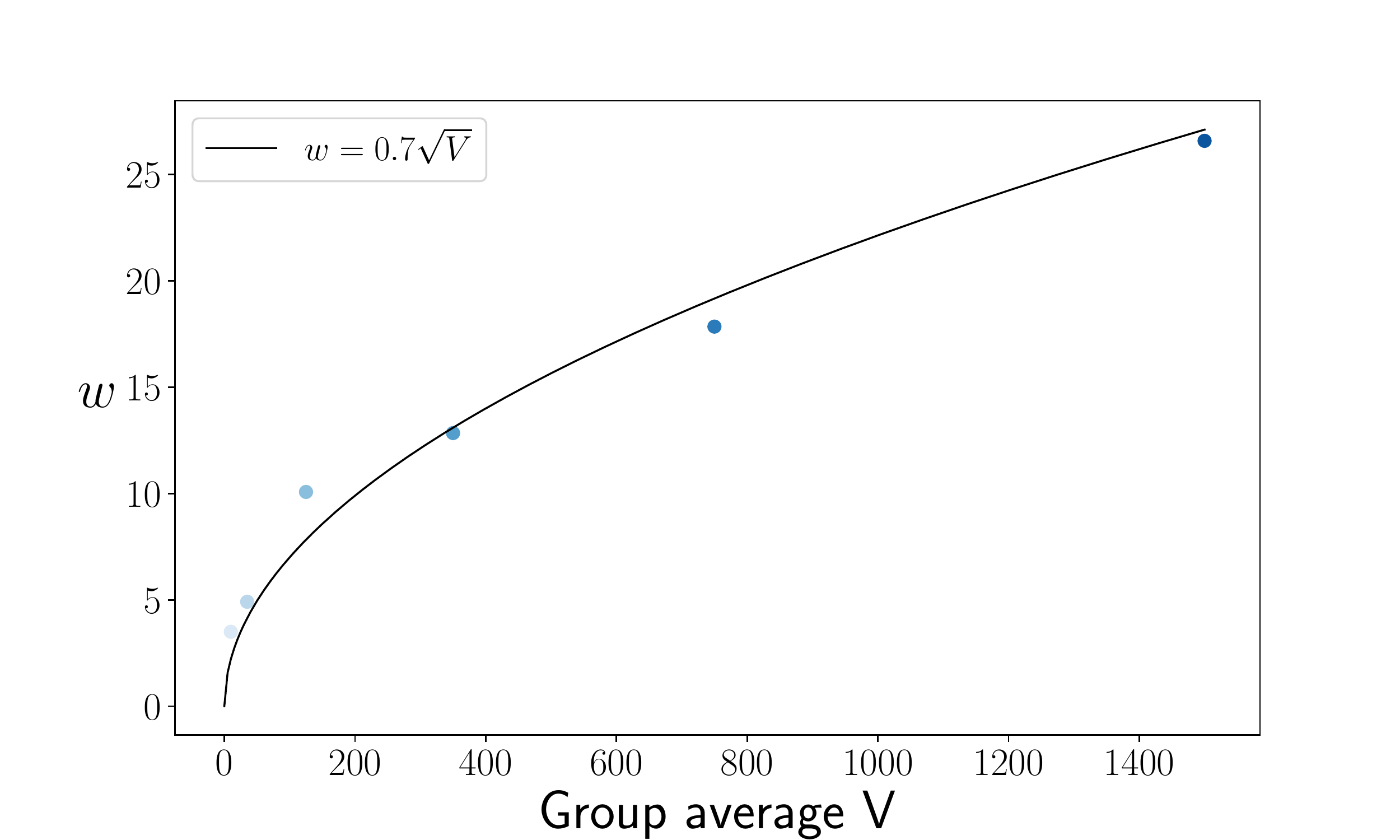}}

\caption{Figure \ref{exp_splitV:a} shows general production curve approximation found with mean squared error minimization in exponential functional family $f_{exp}(\Delta t, w) = 1 - e^{-\frac{\Delta t}{w}}$. Figure  \ref{exp_splitV:b} demonstrates curves found for oil reservoirs groups with various amount of original oil in place. Figure \ref{exp_splitV:c} shows that dependence between original oil in place and $w$ can be approximated with squre root. Identical situation in case of hyperbolic functional family.}
\label{fig:exp_splitV}
\end{figure}

\begin{align}
&f_{hyp}(\Delta t, V, w_0, w_1) = \frac{\Delta t}{\Delta t + w_1 \sqrt{V} + w_0},\\
&f_{exp}(\Delta t, V, w_0, w_1) = 1 - e^{-\frac{\Delta t}{w_1 \sqrt{V} + w_0}}.
\end{align}

Using a training set and any optimization algorithm, we can find an approximation of the general production curve in a functional space. We used gradient descent with Means Squared Error loss. This approximation could be used for oil recovery factor estimation with knowing $dt$, $V$ and $P$. Table \ref{table:extra_features_metrics} lists the oil recovery factor errors metrics on cross-validation for different functional families. As a baseline, the table also shows errors metrics for the simplest oil recovery factor estimation: $\frac{P}{V}$. It is easy to see that the general production curve approach approximation from hyperbolic functional family is much accurate than baseline $\frac{P}{V}$. To enhance the predictive accuracy of machine learning models, we use the oil recovery factor approximations from general production curves as an extra input (this approach is also known as stacking). As potential extra input features we consider  $\frac{P}{V}$, $rf_{exp}(\Delta t, P)$, $rf_{exp}(\Delta t, V, P)$, $rf_{hyp}(\Delta t, P)$, $rf_{hyp}(\Delta t, V, P)$.

\FloatBarrier
\begin{table}[!h]
\begin{center}
\begin{tabular}{|c||c|c|c|c|c|}
\hline
& $\frac{P}{V}$ & $rf_{exp}(\Delta t, P)$ & $rf_{exp}(\Delta t, V, P)$ & $rf_{hyp}(\Delta t, P)$ & $rf_{hyp}(\Delta t, V, P)$ \\
\hline
\hline
MAE  & 10.13 & 9.33 & 8.64 & 8.26 & 7.78\\
$R^2$ & 0.12 & 0.23 & 0.29 & 0.35 & 0.44\\
\hline
\end{tabular}
\caption{Oil recovery factor error metrics using general production curve approximation from different functional families in comparison with simplest baseline $\frac{P}{V}$.}
\label{table:extra_features_metrics}
\end{center}
\end{table}

\begin{figure}[!h]
  \centering
    \includegraphics[width=1.\textwidth]{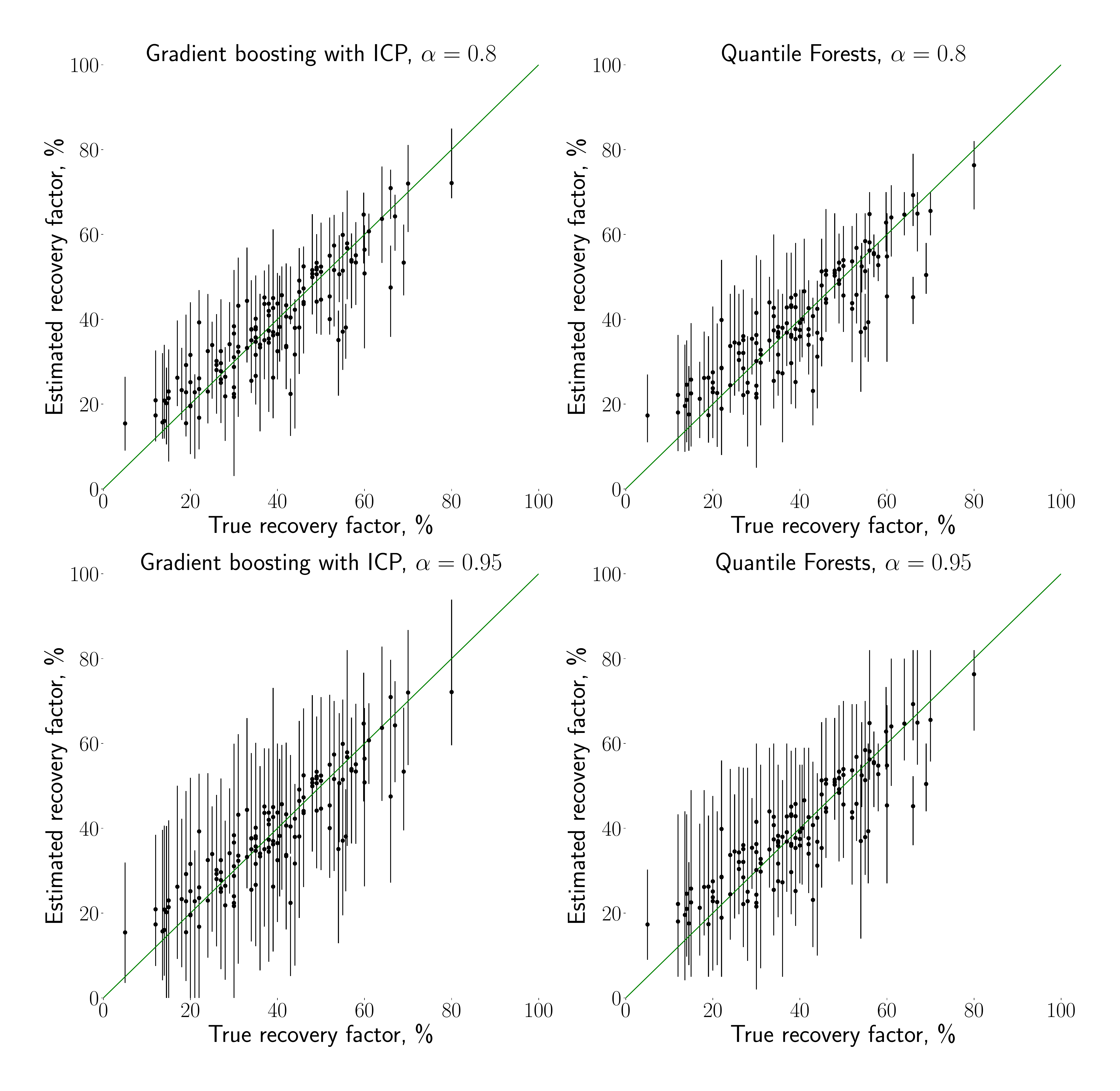}
    \caption{Prediction intervals visualization for 80\% and 95\% confidence levels. Depicted results for randomly picked 25\% points from 20 fold cross-validation.}
    \label{fig:time_dependent_confidence}
\end{figure}

\begin{figure}[!h]
  \centering
    \includegraphics[width=1.\textwidth]{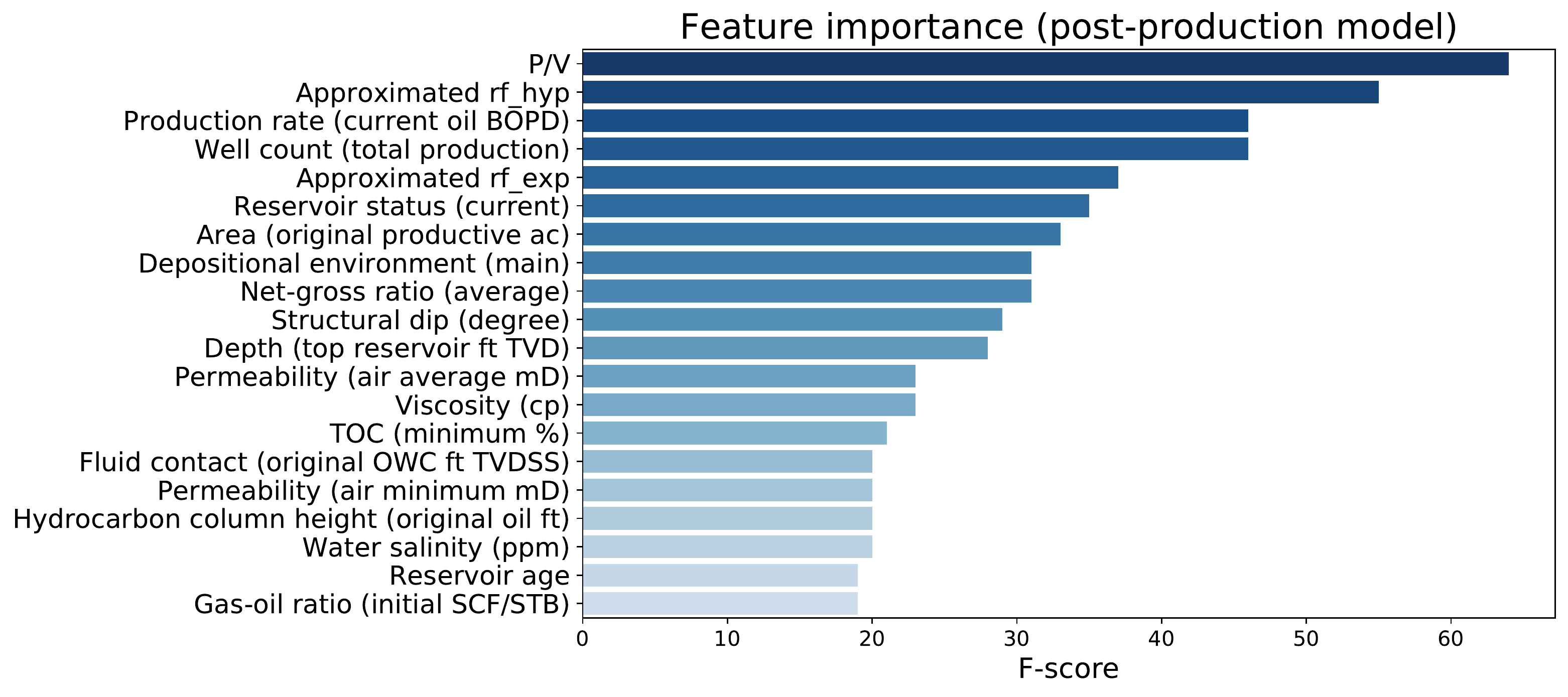}
    \caption{Top 20 features importance for the final Gradient Boosting post-production model. Importance represents as F-score -  the number of times a feature is used to split the data across all trees in ensemble.}
    \label{fig:fi_postprod}
\end{figure}

Table \ref{table:extra_feature_subsets} demonstrates the synergy effect from the combination of general production curves approximation approach and machine learning models. There is a significant improvement in accuracy with using the following extra features subset $\frac{P}{V}$, $rf_{exp}(\Delta t, V, P)$ and $rf_{hyp}(\Delta t, V, P)$. Table \ref{table:dependent_intervals} demonstrates prediction intervals validity and lists its mean width for both Gradient Boosting with ICP and Quantile Regression Forests. Figure \ref{fig:time_dependent_confidence} depicts prediction intervals at 80\% and 95\% confidence levels for both algorithms. Figure \ref{fig:fi_postprod} confirms that features related to production and development are the most important.

\begin{table}[!h]
\begin{center}
\begin{tabular}{|c||c|c|c|c|c|}
\hline
\multicolumn{6}{|c|}{Gradient Boosting} \\
\hline
Extra features & - & $\frac{P}{V}$ & $rf_{exp}(\Delta t, V, P)$ & $rf_{hyp}(\Delta t, V, P)$ & $\frac{P}{V}, rf_{exp, hyp}(\Delta t, V, P)$ \\
\hline
\hline
MAE & 8.56  & 5.13 & 5.29  & 5.08 & \textbf{4.91} \\
$R^2$ & 0.48 & 0.78 & 0.77  & 0.79 & \textbf{0.80}\\
\hline
\hline
\multicolumn{6}{|c|}{Random Forests} \\
\hline
Extra features & - & $\frac{P}{V}$ & $rf_{exp}(\Delta t, V, P)$ & $rf_{hyp}(\Delta t, V, P)$ & $\frac{P}{V}, rf_{exp, hyp}(\Delta t, V, P)$ \\
\hline
\hline
MAE & 9.45  & 5.55 & 5.63  & 5.47 & 5.30 \\
$R^2$ & 0.37 & 0.75 & 0.75  & 0.77 & 0.78\\
\hline
\end{tabular}
\caption{Table lists error metrics calculated on 20 fold cross-validation. It shows effect from adding  $\frac{P}{V}$ (ratio of cumulative oil production to original oil in place), $rf_{exp}(\Delta t, V, P)$ (recovery factor estimation based on general production curve from exponential functional family) and $rf_{hyp}(\Delta t, V, P)$ (recovery factor estimation based on general production curve from hyperbolic functional family). The last column relates to adding subset of all three features.} 
\label{table:extra_feature_subsets}
\end{center}
\end{table}

\begin{table}[!h]
\begin{center}
\begin{tabular}{|c||c|c|c||c|c|c|}
\hline
& \multicolumn{3}{c||}{GB with ICP} & \multicolumn{3}{c|}{QRF} \\
\hline
$\alpha$ & 0.8 & 0.9 & 0.95 & 0.8 & 0.9 & 0.95 \\
\hline
\hline
Mean width  & 18.01 & 24.76 & 32.66 & 17.39 & 22.48 & 26.61\\
Coverage & 0.80 & 0.9 & 0.95 & 0.84 & 0.91 & 0.94\\
\hline
\end{tabular}
\caption{Mean width and coverage rate calculated on 20 fold cross-validation. Both models appears to be valid, since coverage rates are close to corresponding confidence levels.}
\label{table:dependent_intervals}
\end{center}
\end{table}

\FloatBarrier

\section{Discussion}
\label{sec:discussion}

The pre-production model is supposed to be used during reservoir exploration when often all available information is just averaged reservoir characteristics, which can be estimated by measuring the characteristics at several appraisal wells, as well as using seismic exploration data.

The pre-production phase model's accuracy is relatively low for the whole dataset. Perhaps more complete data on the spatial distribution of such characteristics in the reservoir could reduce the prediction error. However, collecting such data requires too many wells. The oil recovery factor is also strongly influenced by the development scheme and its efficiency, as well as the economic situation during development. Thus, one of the main reasons for the low predictive power of the model is the lack of available at exploration phase information. However, the proposed method has the following advantages over traditional ones:

1. Proposed models are general. The algorithms were trained on a representative training set that contains reservoirs from all around the world.

2. The method estimates the prediction uncertainty (predictive intervals), which means that the model can estimate prediction intervals for the corresponding confidence levels. Traditional methods provide only a point estimate, which is not reliable for decision making.

3. The method is computationally cheap (fractions of a second) and does not require any user's special knowledge. Low model's accuracy led us to analyze if there are subset (s) presence for which dependence between parameters and recovery factor stronger than for others.

We conducted a cluster analysis and identified two groups of oil reservoirs. Reservoirs from the first group (cluster) are characterized by less than $200$ million years geological age, predominantly terrigenous deposits, relatively high porosity and permeability. Prediction models demonstrate relatively accurate results for this group and can be used by reservoir experts to assess the potential of the hydrocarbon reservoirs. The best results showed Gradient Boosting with $MAE = 9.57$ and $R^2=0.47$. For the second group, models proved to be less accurate than for the first group. The best metrics shows Gradient Boosting: $MAE = 8.63$ and $R^2 = 0.1$. For both cases, models provide reliable predictive intervals. Based on these facts, we can assume that for carbonate, low-permeability, low-porosity reservoirs, the oil recovery factor's dependence on presented input parameters is much lower. Similar studies have demonstrated more accurate results. However, reservoirs data used in those studies are from a localized geographic area or the result of aggressive filtering \cite{sharma2010classification, aliyuda2019machine}. These papers and the results of the current research lead to a conclusion that there are groups of reservoirs with strong dependency of recovery factor on the reservoir parameters. At the same time, there are a lot of examples where this dependency is very weak.

The post-production phase model gives more accurate predictions due to the fact that the number of input parameters is much greater. It includes development parameters such as well spacing, well count etc. Also, production parameters, such as production rate and cumulative oil production (P) provide information on how efficiently the field is being developed. In particular, the ratio of cumulative oil production (P) and original oil in place (V) gives a close lower bound to oil recovery factor (rf) (Figure \ref{fig:rf_vs_pv}). This lower estimate was further improved with the production curve approximation (Table \ref{table:extra_features_metrics}). The result was used as extra input features for machine learning models (stacking). Figure \ref{fig:fi_postprod} demonstrates that the most important features are related to production and development. We use exponential and hyperbolic functional families to approximate general production curves and use the stacking technique to combine these models with tree-based ensembles. The best model demonstrates the following error metrics calculated on cross-validation: $MAE = 4.91$ and $R^2 = 0.8$. Model also provides reliable predictive intervals. Overall, the model demonstrates predictive power and capability to help experts to optimize development plan as well as to validate results of the hydrodynamic model. 

\section{Conclusion}
\label{sec:conclusion}
In this work, we built and evaluated two tree-based uncertainty quantification models in application to estimating expected ultimate oil recovery factor. We conducted a separate analysis for pre-production phases case and post-production phases.

Model for pre-production phases takes a set of parameters available during reservoir exploration as input. The resulting training set contains 407 oil reservoirs from all around the world, described by 16 time-independent parameters with no more than one missing value. Though the accuracy of the model on the whole training set is low, we identified a group of reservoirs with stronger dependency using clustering analysis. This group of reservoirs are characterized by higher porosity, higher permeability, mostly with terrigenous sediments and with significant difference in the geological age of the rock. This result and several other studies lead us to the conclusion that there are groups of reservoirs with a strong dependency of recovery factor on the reservoir parameters \cite{sharma2010classification, aliyuda2019machine}.

Model for post-production phases takes an extended set of parameters including production and development data. The resulting training set contains 549 oil reservoirs from all around the world, described by 67 parameters. In this case cross-validation metrics show a much higher accuracy. This along with feature importance analysis indicates that the production and development data include a significant quantity of information on oil recovery factor. The application of the production curve approximation approach makes the model even more accurate.

The data-driven technique might be used as a tool for the prompt and objective assessment of reservoir potential due to the richness of the data used for training. It requires much less time and efforts to estimate recovery factor in comparison to existing mature and standard methods. In addition, there is an option to use partial input data for the oil reservoir for assessment. Another advantage of the prediction model is the ability to estimate prediction intervals for the corresponding confidence levels. The trained model generates the recovery factor prediction and calculates the error within a fraction of a second on just a modern office laptop, which is orders of magnitude faster than the most advanced 2D \citep{jin2019deep} and 3D \citep{temirchev2020deep, simonov2018application, temirchev2019reduced} reservoir simulators. These simulators combine differential equations and deep learning techniques. Overall, machine learning has demonstrated its capability to assess the potential of the hydrocarbon reservoirs.

Additional data about different types of reservoirs could allow building more accurate predictive models. Several authors have consider artificial reservoir generation using design of experiment methods and hydrocarbon reservoir simulators. A notable examples of this approach are presented in \cite{naderi2016nonlinear} and \cite{panja2018application}. The future research will consider ways to increase training set size using hydrodynamic simulators or its surrogate models \citep{temirchev2019reduced}.

\newpage

\section*{Acknowledgement}
The work of Evgeny Burnaev in Sections was supported by Ministry of Science and Higher Education grant No. 075-10-2021-068
\newpage

\bibliography{bibfile}

\begin{thebibliography}{35}
\expandafter\ifx\csname natexlab\endcsname\relax\def\natexlab#1{#1}\fi
\providecommand{\url}[1]{\texttt{#1}}
\providecommand{\href}[2]{#2}
\providecommand{\path}[1]{#1}
\providecommand{\DOIprefix}{doi:}
\providecommand{\ArXivprefix}{arXiv:}
\providecommand{\URLprefix}{URL: }
\providecommand{\Pubmedprefix}{pmid:}
\providecommand{\doi}[1]{\href{http://dx.doi.org/#1}{\path{#1}}}
\providecommand{\Pubmed}[1]{\href{pmid:#1}{\path{#1}}}
\providecommand{\bibinfo}[2]{#2}
\ifx\xfnm\relax \def\xfnm[#1]{\unskip,\space#1}\fi
\bibitem[{Rui et~al.(2017)Rui, Lu, Zhang, Guo, Ling, Zhang, and
  Patil}]{rui2017quantitative}
\bibinfo{author}{Z.~Rui}, \bibinfo{author}{J.~Lu}, \bibinfo{author}{Z.~Zhang},
  \bibinfo{author}{R.~Guo}, \bibinfo{author}{K.~Ling},
  \bibinfo{author}{R.~Zhang}, \bibinfo{author}{S.~Patil},
\newblock \bibinfo{title}{A quantitative oil and gas reservoir evaluation
  system for development},
\newblock \bibinfo{journal}{Journal of Natural Gas Science and Engineering}
  \bibinfo{volume}{42} (\bibinfo{year}{2017}) \bibinfo{pages}{31--39}.
  \DOIprefix\doi{https://doi.org/10.1016/j.jngse.2017.02.026}.
\bibitem[{Demirmen et~al.(2007)}]{demirmen2007reserves}
\bibinfo{author}{F.~Demirmen}, et~al.,
\newblock \bibinfo{title}{Reserves estimation: the challenge for the industry},
\newblock \bibinfo{journal}{Journal of Petroleum Technology}
  \bibinfo{volume}{59} (\bibinfo{year}{2007}) \bibinfo{pages}{80--89}.
\bibitem[{Li et~al.(2020)Li, Yu, Cao, Tian, and Cheng}]{li2020applications}
\bibinfo{author}{H.~Li}, \bibinfo{author}{H.~Yu}, \bibinfo{author}{N.~Cao},
  \bibinfo{author}{H.~Tian}, \bibinfo{author}{S.~Cheng},
\newblock \bibinfo{title}{Applications of artificial intelligence in oil and
  gas development},
\newblock \bibinfo{journal}{Archives of Computational Methods in Engineering}
  (\bibinfo{year}{2020}) \bibinfo{pages}{1--13}.
\bibitem[{Guthrie et~al.(1955)Guthrie, Greenberger et~al.}]{guthrie1955use}
\bibinfo{author}{R.~Guthrie}, \bibinfo{author}{M.~H. Greenberger}, et~al.,
\newblock \bibinfo{title}{The use of multiple-correlation analyses for
  interpreting petroleum-engineering data},
\newblock in: \bibinfo{booktitle}{Drilling and Production Practice},
  \bibinfo{organization}{American Petroleum Institute}, \bibinfo{year}{1955}.
\bibitem[{Arps et~al.(1967)Arps, Brons, Van~Everdingen, Buchwald, and
  Smith}]{arps1967statistical}
\bibinfo{author}{J.~Arps}, \bibinfo{author}{F.~Brons},
  \bibinfo{author}{A.~Van~Everdingen}, \bibinfo{author}{R.~Buchwald},
  \bibinfo{author}{A.~Smith},
\newblock \bibinfo{title}{A statistical study of recovery efficiency},
\newblock \bibinfo{journal}{Bull. D} \bibinfo{volume}{14}
  (\bibinfo{year}{1967}).
\bibitem[{Sharma et~al.(2010)Sharma, Srinivasan, Lake
  et~al.}]{sharma2010classification}
\bibinfo{author}{A.~Sharma}, \bibinfo{author}{S.~Srinivasan},
  \bibinfo{author}{L.~W. Lake}, et~al.,
\newblock \bibinfo{title}{Classification of oil and gas reservoirs based on
  recovery factor: a data-mining approach},
\newblock in: \bibinfo{booktitle}{SPE Annual Technical Conference and
  Exhibition}, \bibinfo{organization}{Society of Petroleum Engineers},
  \bibinfo{year}{2010}. \DOIprefix\doi{http://dx.doi.org/10.2118/130257-MS}.
\bibitem[{Mahmoud et~al.(2019)Mahmoud, Elkatatny, Chen, and
  Abdulraheem}]{mahmoud2019estimation}
\bibinfo{author}{A.~A. Mahmoud}, \bibinfo{author}{S.~Elkatatny},
  \bibinfo{author}{W.~Chen}, \bibinfo{author}{A.~Abdulraheem},
\newblock \bibinfo{title}{Estimation of oil recovery factor for water drive
  sandy reservoirs through applications of artificial intelligence},
\newblock \bibinfo{journal}{Energies} \bibinfo{volume}{12}
  (\bibinfo{year}{2019}) \bibinfo{pages}{3671}.
\bibitem[{Han and Bian(2018)}]{han2018hybrid}
\bibinfo{author}{B.~Han}, \bibinfo{author}{X.~Bian},
\newblock \bibinfo{title}{A hybrid pso-svm-based model for determination of oil
  recovery factor in the low-permeability reservoir},
\newblock \bibinfo{journal}{Petroleum} \bibinfo{volume}{4}
  (\bibinfo{year}{2018}) \bibinfo{pages}{43--49}.
\bibitem[{Aliyuda and Howell(2019)}]{aliyuda2019machine}
\bibinfo{author}{K.~Aliyuda}, \bibinfo{author}{J.~Howell},
\newblock \bibinfo{title}{Machine-learning algorithm for estimating
  oil-recovery factor using a combination of engineering and stratigraphic
  dependent parameters},
\newblock \bibinfo{journal}{Interpretation} \bibinfo{volume}{7}
  (\bibinfo{year}{2019}) \bibinfo{pages}{SE151--SE159}.
  \DOIprefix\doi{https://doi.org/10.1190/INT-2018-0211.1}.
\bibitem[{Belyaev et~al.(2016)Belyaev, Burnaev, Kapushev, Panov, Prikhodko,
  Vetrov, and Yarotsky}]{GTApprox2016}
\bibinfo{author}{M.~Belyaev}, \bibinfo{author}{E.~Burnaev},
  \bibinfo{author}{E.~Kapushev}, \bibinfo{author}{M.~Panov},
  \bibinfo{author}{P.~Prikhodko}, \bibinfo{author}{D.~Vetrov},
  \bibinfo{author}{D.~Yarotsky},
\newblock \bibinfo{title}{Gtapprox: Surrogate modeling for industrial design},
\newblock \bibinfo{journal}{Advances in Engineering Software}
  \bibinfo{volume}{102} (\bibinfo{year}{2016}) \bibinfo{pages}{29 -- 39}.
  \DOIprefix\doi{https://doi.org/10.1016/j.advengsoft.2016.09.001}.
\bibitem[{tor(2016)}]{toris}
\bibinfo{title}{Toris: An integrated decision support system for petroleum e\&p
  policy evaluation [dataset]},
  \bibinfo{howpublished}{\url{https://data.wu.ac.at/schema/edx_netl_doe_gov/MDBkMzNmM2YtOGQzYi00MWQ0LTkyZmYtZDg0MDgzZjVjODdk}},
  \bibinfo{year}{2016}.
\bibitem[{Burnaev and Vovk(2014)}]{VovkConformal2014}
\bibinfo{author}{E.~Burnaev}, \bibinfo{author}{V.~Vovk},
\newblock \bibinfo{title}{Efficiency of conformalized ridge regression},
\newblock in: \bibinfo{editor}{M.~F. Balcan}, \bibinfo{editor}{V.~Feldman},
  \bibinfo{editor}{C.~Szepesvári} (Eds.), \bibinfo{booktitle}{Proceedings of
  The 27th Conference on Learning Theory}, volume~\bibinfo{volume}{35} of
  \textit{\bibinfo{series}{Proceedings of Machine Learning Research}},
  \bibinfo{year}{2014}, pp. \bibinfo{pages}{605--622}.
\bibitem[{Roy and Larocque(2012)}]{roy2012robustness}
\bibinfo{author}{M.-H. Roy}, \bibinfo{author}{D.~Larocque},
\newblock \bibinfo{title}{Robustness of random forests for regression},
\newblock \bibinfo{journal}{Journal of Nonparametric Statistics}
  \bibinfo{volume}{24} (\bibinfo{year}{2012}) \bibinfo{pages}{993--1006}.
\bibitem[{G{\'o}mez-R{\'\i}os et~al.(2017)G{\'o}mez-R{\'\i}os, Luengo, and
  Herrera}]{gomez2017study}
\bibinfo{author}{A.~G{\'o}mez-R{\'\i}os}, \bibinfo{author}{J.~Luengo},
  \bibinfo{author}{F.~Herrera},
\newblock \bibinfo{title}{A study on the noise label influence in boosting
  algorithms: Adaboost, gbm and xgboost},
\newblock in: \bibinfo{booktitle}{International Conference on Hybrid Artificial
  Intelligence Systems}, \bibinfo{organization}{Springer},
  \bibinfo{year}{2017}, pp. \bibinfo{pages}{268--280}.
\bibitem[{Breiman(2001)}]{breiman2001random}
\bibinfo{author}{L.~Breiman},
\newblock \bibinfo{title}{Random forests},
\newblock \bibinfo{journal}{Machine learning} \bibinfo{volume}{45}
  (\bibinfo{year}{2001}) \bibinfo{pages}{5--32}.
  \DOIprefix\doi{https://doi.org/10.1023/A:1010933404324}.
\bibitem[{Meinshausen(2006)}]{meinshausen2006quantile}
\bibinfo{author}{N.~Meinshausen},
\newblock \bibinfo{title}{Quantile regression forests},
\newblock \bibinfo{journal}{Journal of Machine Learning Research}
  \bibinfo{volume}{7} (\bibinfo{year}{2006}) \bibinfo{pages}{983--999}.
\bibitem[{Friedman(2001)}]{friedman2001greedy}
\bibinfo{author}{J.~H. Friedman},
\newblock \bibinfo{title}{Greedy function approximation: a gradient boosting
  machine},
\newblock \bibinfo{journal}{Annals of statistics}  (\bibinfo{year}{2001})
  \bibinfo{pages}{1189--1232}.
  \DOIprefix\doi{https://doi.org/10.1214/aos/1013203451}.
\bibitem[{Vovk et~al.(2005)Vovk, Gammerman, and Shafer}]{vovk2005algorithmic}
\bibinfo{author}{V.~Vovk}, \bibinfo{author}{A.~Gammerman},
  \bibinfo{author}{G.~Shafer}, \bibinfo{title}{Algorithmic learning in a random
  world}, \bibinfo{publisher}{Springer Science \& Business Media},
  \bibinfo{year}{2005}. \DOIprefix\doi{https://doi.org/10.1007/b106715}.
\bibitem[{Burnaev and Nazarov(2016)}]{ConformalKRR2016}
\bibinfo{author}{E.~Burnaev}, \bibinfo{author}{I.~Nazarov},
\newblock \bibinfo{title}{Conformalized kernel ridge regression},
\newblock in: \bibinfo{booktitle}{2016 15th IEEE International Conference on
  Machine Learning and Applications (ICMLA)}, \bibinfo{year}{2016}, pp.
  \bibinfo{pages}{45--52}. \DOIprefix\doi{10.1109/ICMLA.2016.0017}.
\bibitem[{Hartigan and Wong(1979)}]{hartigan1979algorithm}
\bibinfo{author}{J.~A. Hartigan}, \bibinfo{author}{M.~A. Wong},
\newblock \bibinfo{title}{Algorithm as 136: A k-means clustering algorithm},
\newblock \bibinfo{journal}{Journal of the Royal Statistical Society. Series C
  (Applied Statistics)} \bibinfo{volume}{28} (\bibinfo{year}{1979})
  \bibinfo{pages}{100--108}. \DOIprefix\doi{http://dx.doi.org/10.2307/2346830}.
\bibitem[{Jain(2010)}]{jain2010data}
\bibinfo{author}{A.~K. Jain},
\newblock \bibinfo{title}{Data clustering: 50 years beyond k-means},
\newblock \bibinfo{journal}{Pattern recognition letters} \bibinfo{volume}{31}
  (\bibinfo{year}{2010}) \bibinfo{pages}{651--666}.
  \DOIprefix\doi{http://dx.doi.org/10.1016/j.patrec.2009.09.011}.
\bibitem[{Arthur and Vassilvitskii(2007)}]{arthur2007k}
\bibinfo{author}{D.~Arthur}, \bibinfo{author}{S.~Vassilvitskii},
\newblock \bibinfo{title}{k-means++: The advantages of careful seeding},
\newblock in: \bibinfo{booktitle}{Proceedings of the eighteenth annual ACM-SIAM
  symposium on Discrete algorithms}, \bibinfo{organization}{Society for
  Industrial and Applied Mathematics}, \bibinfo{year}{2007}, pp.
  \bibinfo{pages}{1027--1035}.
\bibitem[{Li et~al.(2017)Li, Cerise, Yang, and Han}]{li2017application}
\bibinfo{author}{W.~Li}, \bibinfo{author}{J.~E. Cerise},
  \bibinfo{author}{Y.~Yang}, \bibinfo{author}{H.~Han},
\newblock \bibinfo{title}{Application of t-sne to human genetic data},
\newblock \bibinfo{journal}{Journal of bioinformatics and computational
  biology} \bibinfo{volume}{15} (\bibinfo{year}{2017})
  \bibinfo{pages}{1750017}.
  \DOIprefix\doi{https://doi.org/10.1142/S0219720017500172}.
\bibitem[{Maaten and Hinton(2008)}]{maaten2008visualizing}
\bibinfo{author}{L.~v.~d. Maaten}, \bibinfo{author}{G.~Hinton},
\newblock \bibinfo{title}{Visualizing data using t-sne},
\newblock \bibinfo{journal}{Journal of machine learning research}
  \bibinfo{volume}{9} (\bibinfo{year}{2008}) \bibinfo{pages}{2579--2605}.
\bibitem[{Twala et~al.(2008)Twala, Jones, and Hand}]{twala2008good}
\bibinfo{author}{B.~Twala}, \bibinfo{author}{M.~Jones}, \bibinfo{author}{D.~J.
  Hand},
\newblock \bibinfo{title}{Good methods for coping with missing data in decision
  trees},
\newblock \bibinfo{journal}{Pattern Recognition Letters} \bibinfo{volume}{29}
  (\bibinfo{year}{2008}) \bibinfo{pages}{950--956}.
\bibitem[{Orlov and Koroteev(2019)}]{orlov2019advanced}
\bibinfo{author}{D.~Orlov}, \bibinfo{author}{D.~Koroteev},
\newblock \bibinfo{title}{Advanced analytics of self-colmatation in terrigenous
  oil reservoirs},
\newblock \bibinfo{journal}{Journal of Petroleum Science and Engineering}
  \bibinfo{volume}{182} (\bibinfo{year}{2019}) \bibinfo{pages}{106306}.
\bibitem[{Erofeev et~al.(2019)Erofeev, Orlov, Ryzhov, and
  Koroteev}]{erofeev2019prediction}
\bibinfo{author}{A.~Erofeev}, \bibinfo{author}{D.~Orlov},
  \bibinfo{author}{A.~Ryzhov}, \bibinfo{author}{D.~Koroteev},
\newblock \bibinfo{title}{Prediction of porosity and permeability alteration
  based on machine learning algorithms},
\newblock \bibinfo{journal}{Transport in Porous Media} \bibinfo{volume}{128}
  (\bibinfo{year}{2019}) \bibinfo{pages}{677--700}.
\bibitem[{Kotsiantis(2013)}]{kotsiantis2013decision}
\bibinfo{author}{S.~B. Kotsiantis},
\newblock \bibinfo{title}{Decision trees: a recent overview},
\newblock \bibinfo{journal}{Artificial Intelligence Review}
  \bibinfo{volume}{39} (\bibinfo{year}{2013}) \bibinfo{pages}{261--283}.
\bibitem[{Fetkovich et~al.(1996)Fetkovich, Fetkovich, Fetkovich
  et~al.}]{fetkovich1996useful}
\bibinfo{author}{M.~Fetkovich}, \bibinfo{author}{E.~Fetkovich},
  \bibinfo{author}{M.~Fetkovich}, et~al.,
\newblock \bibinfo{title}{Useful concepts for decline curve forecasting,
  reserve estimation, and analysis},
\newblock \bibinfo{journal}{SPE Reservoir Engineering} \bibinfo{volume}{11}
  (\bibinfo{year}{1996}) \bibinfo{pages}{13--22}.
  \DOIprefix\doi{http://dx.doi.org/10.2118/28628-PA}.
\bibitem[{Jin et~al.(2019)Jin, Liu, and Durlofsky}]{jin2019deep}
\bibinfo{author}{Z.~L. Jin}, \bibinfo{author}{Y.~Liu}, \bibinfo{author}{L.~J.
  Durlofsky},
\newblock \bibinfo{title}{Deep-learning-based reduced-order modeling for
  subsurface flow simulation},
\newblock \bibinfo{journal}{arXiv preprint arXiv:1906.03729}
  (\bibinfo{year}{2019}).
\bibitem[{Temirchev et~al.(2020)Temirchev, Simonov, Kostoev, Burnaev,
  Oseledets, Akhmetov, Margarit, Sitnikov, and Koroteev}]{temirchev2020deep}
\bibinfo{author}{P.~Temirchev}, \bibinfo{author}{M.~Simonov},
  \bibinfo{author}{R.~Kostoev}, \bibinfo{author}{E.~Burnaev},
  \bibinfo{author}{I.~Oseledets}, \bibinfo{author}{A.~Akhmetov},
  \bibinfo{author}{A.~Margarit}, \bibinfo{author}{A.~Sitnikov},
  \bibinfo{author}{D.~Koroteev},
\newblock \bibinfo{title}{Deep neural networks predicting oil movement in a
  development unit},
\newblock \bibinfo{journal}{Journal of Petroleum Science and Engineering}
  \bibinfo{volume}{184} (\bibinfo{year}{2020}) \bibinfo{pages}{106513}.
\bibitem[{Simonov et~al.(2018)Simonov, Akhmetov, Temirchev, Koroteev, Kostoev,
  Burnaev, Oseledets et~al.}]{simonov2018application}
\bibinfo{author}{M.~Simonov}, \bibinfo{author}{A.~Akhmetov},
  \bibinfo{author}{P.~Temirchev}, \bibinfo{author}{D.~Koroteev},
  \bibinfo{author}{R.~Kostoev}, \bibinfo{author}{E.~Burnaev},
  \bibinfo{author}{I.~Oseledets}, et~al.,
\newblock \bibinfo{title}{Application of machine learning technologies for
  rapid 3d modelling of inflow to the well in the development system},
\newblock in: \bibinfo{booktitle}{SPE Russian Petroleum Technology Conference},
  \bibinfo{organization}{Society of Petroleum Engineers}, \bibinfo{year}{2018}.
\bibitem[{Temirchev et~al.(2019)Temirchev, Gubanova, Kostoev, Gryzlov,
  Voloskov, Koroteev, Simonov, Akhmetov, Margarit, Ershov
  et~al.}]{temirchev2019reduced}
\bibinfo{author}{P.~Temirchev}, \bibinfo{author}{A.~Gubanova},
  \bibinfo{author}{R.~Kostoev}, \bibinfo{author}{A.~Gryzlov},
  \bibinfo{author}{D.~Voloskov}, \bibinfo{author}{D.~Koroteev},
  \bibinfo{author}{M.~Simonov}, \bibinfo{author}{A.~Akhmetov},
  \bibinfo{author}{A.~Margarit}, \bibinfo{author}{A.~Ershov}, et~al.,
\newblock \bibinfo{title}{Reduced order reservoir simulation with
  neural-network based hybrid model},
\newblock in: \bibinfo{booktitle}{SPE Russian Petroleum Technology Conference},
  \bibinfo{organization}{Society of Petroleum Engineers}, \bibinfo{year}{2019}.
\bibitem[{Naderi and Khamehchi(2016)}]{naderi2016nonlinear}
\bibinfo{author}{M.~Naderi}, \bibinfo{author}{E.~Khamehchi},
\newblock \bibinfo{title}{Nonlinear risk optimization approach to water drive
  gas reservoir production optimization using doe and artificial intelligence},
\newblock \bibinfo{journal}{Journal of Natural Gas Science and Engineering}
  \bibinfo{volume}{31} (\bibinfo{year}{2016}) \bibinfo{pages}{575--584}.
\bibitem[{Panja et~al.(2018)Panja, Velasco, Pathak, and
  Deo}]{panja2018application}
\bibinfo{author}{P.~Panja}, \bibinfo{author}{R.~Velasco},
  \bibinfo{author}{M.~Pathak}, \bibinfo{author}{M.~Deo},
\newblock \bibinfo{title}{Application of artificial intelligence to forecast
  hydrocarbon production from shales},
\newblock \bibinfo{journal}{Petroleum} \bibinfo{volume}{4}
  (\bibinfo{year}{2018}) \bibinfo{pages}{75--89}.

\end{thebibliography}

\end{document}